\documentstyle[12pt,aasms4]{article}
\def\MG{MG~1131+0456~}

\def\arcs{\ifmmode {^{\scriptscriptstyle\prime\prime}}
          \else $^{\scriptscriptstyle\prime\prime}$\fi}
\def\arcm{\ifmmode {^{\scriptscriptstyle\prime}}
          \else $^{\scriptscriptstyle\prime}$\fi}
\newdimen\sa  \newdimen\sb
\def\parcs{\sa=.07em \sb=.03em
     \ifmmode $\rlap{.}$^{\scriptscriptstyle\prime\kern -\sb\prime}$\kern -\sa$
     \else \rlap{.}$^{\scriptscriptstyle\prime\kern -\sb\prime}$\kern -\sa\fi}
\def\parcm{\sa=.08em \sb=.03em
     \ifmmode $\rlap{.}\kern\sa$^{\scriptscriptstyle\prime}$\kern-\sb$
     \else \rlap{.}\kern\sa$^{\scriptscriptstyle\prime}$\kern-\sb\fi}
\def\pdeg{\ifmmode $\setbox0=\hbox{$^{\circ}$}\rlap{\hskip.11\wd0 .}$^{\circ}
          \else \setbox0=\hbox{$^{\circ}$}\rlap{\hskip.11\wd0 .}$^{\circ}$\fi}
\def\gtorder{\mathrel{\raise.3ex\hbox{$>$}\mkern-14mu
             \lower0.6ex\hbox{$\sim$}}}
\def\ltorder{\mathrel{\raise.3ex\hbox{$<$}\mkern-14mu
             \lower0.6ex\hbox{$\sim$}}}
\def\eg{{\it e.g.~}}

\newcommand{\kms}{\mbox{ km~s$^{-1}$}}

\slugcomment{Submitted to Astrophysical Journal}

\input psfig
\lefthead{Kochanek et al.}
\righthead{}

\begin{document}

\title{The Infrared Einstein Ring in the Gravitational Lens \MG \\
   and the Death of the Dusty Lens Hypothesis\footnote{Based on Observations made with the NASA/ESA
  Hubble Space Telescope, obtained at the Space Telescope Science Institute, which is operated
  by AURA, Inc., under NASA contract NAS 5-26555. } }

\vskip 2truecm

\author{C. S. Kochanek$^{(a)}$,} 

\author{E. E. Falco$^{(a)}$, C. D. Impey$^{(b)}$, J. Leh\'ar$^{(a)}$, B. A. McLeod$^{(a)}$}

\author{H.-W. Rix$^{(b)}$, C. R. Keeton$^{(b)}$ and C.Y. Peng$^{(b)}$}

\affil{$^{(a)}$ Harvard-Smithsonian Center for Astrophysics, 60 Garden St., Cambridge,
	MA 02138}
\affil{email: ckochanek, efalco, jlehar, bmcleod@cfa.harvard.edu}
\affil{$^{(b)}$Steward Observatory, University of Arizona, Tucson, AZ 85721}
\affil{email: cimpey, rix, ckeeton, cyp@as.arizona.edu}

\begin{abstract}
We have obtained and modeled new NICMOS images of the lens system \MG, which show that
its lens galaxy is an H$=18.6$ mag, transparent, early-type galaxy
at a redshift of $z_l \simeq 0.85$; it has a major axis effective radius 
$R_e=0\parcs68\pm0\parcs05$, projected axis ratio $b/a=0.77\pm0.02$, and major axis 
PA$=60^\circ \pm 2^\circ$.  The lens is the brightest member of 
a group of seven galaxies with similar R--I and I--H colors, and the two closest group members
produce sufficient tidal perturbations  to explain
the ring morphology.  The host galaxy of the \MG source is a 
$z_s \gtorder 2$ ERO (``extremely red object'') which is lensed 
into optical and infrared rings of dramatically different morphologies. These
differences imply a strongly wavelength-dependent source morphology 
that could  be explained by embedding the host in a larger, dusty disk.  At
1.6$\mu$m (H), the
ring is spectacularly luminous, with a total observed flux of H$=17.4$ mag and a 
de-magnified flux of $19.3$ mag, corresponding to a 1--2$L_*$
galaxy at the probable source redshift of $z_s \gtorder 2$.   Thus, it is
primarily the stellar emission of the radio source host galaxy that produces the 
overall colors of two of the reddest radio lenses, \MG and B~1938+666,
aided by the suppression of optical AGN emission by dust in
the source galaxy.  The dusty lens hypothesis --- that many massive early-type
galaxies with $0.2 \ltorder z_l \ltorder 1.0$ have large, uniform dust opacities
--- is ruled out.
\end{abstract}


\section{INTRODUCTION}

The lens galaxies of the $\sim30$ gravitational lenses with
$\Delta\theta\gtorder1\parcs0$ constitute
a sample of massive galaxies selected from diverse but generally low
density environments in the redshift range $0.2 \ltorder z_l \ltorder 1.0$.  
As such, they complement the evolutionary studies of
galaxies in rich clusters across the same redshift range.
  Stanford et al. (1995, 1998) and Ellis et al. (1997)
have shown that morphologically identified early-type galaxies in rich clusters
have colors that match passively evolving stellar population models even at
$z\simeq 1$.  These galaxies obey fundamental-plane relations, and the directly
measured evolution of the mass-to-light ratio ($M/L$) is well matched to the
changes expected from passive evolution (e.g.  Kelson et al. 1997, van Dokkum et al. 1998).
Keeton, Kochanek \& Falco (1998) showed that the gravitational lens galaxies show 
the same passive evolution in both color and $M/L$ as their counterparts
in rich clusters.  While there are fewer lens galaxies, they have the great 
advantage of precisely measured individual masses at any redshift without
the complications of interpreting stellar dynamics.  

Studies of red galaxies with $z\gtorder 1$ are more difficult due
to the lack of identified clusters and the difficulty in obtaining spectra of
red objects (as compared to blue, star-forming, emission line objects).  In
particular, the properties of the ERO (extremely red object) population
with R--K$\gtorder 6$ mag are poorly understood. For example, LBDS~53W091
at $z=1.55$ (Dunlop et al. 1995, Spinrad et al. 1997) is believed to be
a nearly dust free, early-type galaxy whose old stellar population sets
strong limits on the epoch of star formation and the age of the 
universe at that epoch.
On the other hand, Graham \& Dey (1996) argue that the less radio luminous galaxy 
HR~10 at $z=1.44$ is an intense starburst which is heavily obscured by dust.
Quasars and radio sources also show a wide range of optical colors.  
Flat-spectrum radio sources range from very blue quasars to colors as
red as those of the EROs (Webster et al. 1995), with the optically selected 
quasars concentrated on the blue tail of the distribution. The red colors
can be due to the extinction in the host galaxy (e.g. Wills et al. 1992),
extinction in intervening galaxies or protogalaxies (e.g. Heisler \& Ostriker 1988,
Fall \& Pei 1993), the continuation of steep synchrotron emission from the
compact radio source (e.g. Impey \& Neugebauer 1988), or 
the stellar light of the host galaxy. The last possibility was assumed by 
Dunlop et al. (1996) when they selected LBDS~53W091. 

Gravitational lens systems also offer a unique opportunity to probe the 
dust content of distant galaxies,
because the lensed images can be used to estimate
the extinction differences on paths separated by 1--10 kpc 
at typical redshifts (e.g. Nadeau et al. 1991).
Further, differences between the statistics of radio
and optically selected lens samples can be used to estimate the mean extinction of
the population (Falco, Kochanek \& Mu\~noz 1998).  In this approach there is no need
to untangle the amount and distribution of dust from stellar population effects.
Surveys of radio-selected lenses by
Annis \& Luppino (1993) and Malhotra, Rhoads \& Turner (1997) showed that
a significant number had very red optical to infrared colors, particularly
the radio-selected lenses MG~0414+0534  (Hewitt et al. 1992, Lawrence et al. 1995,
Annis \& Luppino 1993), \MG (Hewitt et al. 1988, Annis 1992, Larkin et al. 1994), 
and B~1938+666 (King et al. 1997, Rhoads, Malhotra, \& Kundic 1996).  With typical R--K colors of 6--7 mag, 
these three lenses lie near the extreme red edge of the color distribution found by 
Webster et al. (1995) and have colors typical of the EROs.
Of the three, only MG~0414+0534 provides direct evidence for dust: the
optical images are dominated by four point-like images (Hewitt et al. 1992) of
a highly reddened ($A_V\sim 7$), $z_s=2.64$ quasar identified by its broad H$\alpha$ line 
in the infrared (Lawrence et al. 1995).  In the other two cases the dust is
inferred from the red colors by assuming that the source should have an
intrinsically blue quasar spectrum due to the presence of the radio 
source, but there is no direct support from a well-resolved image or an
analyzable spectrum.  The two surveys of lens colors reached
quite different conclusions.  Malhotra et al. (1997) advocated dusty lenses as
the origin of the red colors, while Annis \& Luppino (1993) advocated reddening 
in the source combined with stellar emission from the host galaxy.

With the advent of WFPC2 and NICMOS it became possible to image directly 
the lenses and study the morphology and colors of the host galaxies.
Emission by the host galaxy is seen in the 
optical for \MG (Hammer et al. 1991, Keeton et al. 1998),
MG~0414+0534 (Falco et al. 1997), BRI~0952--0115 (Leh\'ar et al. 1998), 
B~1600+434 (Keeton et al. 1998), B~1608+656 (Fassnacht et al. 1996)
and FSC~10214+4724 (Eisenhardt et al. 1996) and in the infrared for B~1938+666 (King et al. 1998)
PG~1115+080 (Impey et al. 1998), 
HE~1104--1805 (Leh\'ar et al. 1998) and H~1413+117 (McLeod et al. 1998b).
Near-infrared data have the natural advantage for studying host galaxies,
because the H filter measures flux emitted longward of the 4000\AA~ break for
all  source redshifts $z_s \ltorder 3$. In contrast, the I-band records only the 
relatively faint rest-UV flux for redshifts $z_s \gtorder 1$.  

\MG  was the first Einstein 
ring lens to be discovered (Hewitt et al. 1988).  It consists of
two images of a radio core and an Einstein ring created by the radio
jet crossing the astroid caustic of the lens.  The ring is very
elliptical with a major axis diameter of $2\parcs05$ between the
radio cores.  Chen \& Hewitt (1993) and Hewitt, Chen \& Messier  (1995) produced 
greatly improved radio images and found that the cores are weakly
variable.  The optical counterpart to the system is faint ($m_R \simeq 22$ mag)
and red.  Hammer et al. (1991) suggested the lens was an early-type galaxy at 
$z_l \simeq 0.85$ and that there was an optical Einstein ring surrounding the 
central lens galaxy.  Infrared imaging by Annis (1992) and Larkin et al. (1994) 
showed that the counterparts of the radio cores are very red and that 
there is a concentration of galaxies projected within 20$^{\prime\prime}$ 
of the lens. Larkin et al. (1994) attributed the red source color to extinction
in the lens galaxy, while Annis \& Luppino (1993) interpreted it as 
 source galaxy flux.  Kochanek et al. (1989) and Chen, Kochanek \& Hewitt (1995)
modeled the radio emission to reconstruct the lens mass distribution.   
Chen et al. (1995) found that the shear axis of the best fit models
matched the orientation of the tidal shear produced by the two nearby galaxies
found by Annis (1992),  and Keeton et al. (1998) found that the shear axis
of the models was misaligned with respect to the major axis of the
lens galaxy.  

In this paper we present new H band images of MG~1131+0456, obtained
with NICMOS as part of the Center for Astrophysics/Arizona Space Telescope
Lens Survey (CASTLES). These images permit
a detailed study of the lensing galaxy, the source galaxy,
 the dust content of both lens and source, and allow us to estimate
the redshifts of a galaxy group projected nearby.
We start by describing the data (\S2.1) and the
spectrophotometric models we use for interpretation (\S2.2).
Next, we describe qualitatively the wavelength dependent structure
of the source (\S2.3) and follow that with a detailed discussion of the
lens galaxy (\S2.4) and the surrounding group (\S2.5). Next
we derive a new lens model for the system (\S3.1) and
then use the model to reconstruct the intrinsic source structure 
(\S3.2).  In \S4 we discuss the extinction in \MG and
its consequences for hypotheses about the role of dust in
lens systems.  Finally, we summarize our findings in \S5. 

\section{Observations and Data Analysis}

\subsection {HST Photometry}

Using the NIC2 camera on HST
we observed \MG through the F160W filter,
which corresponds roughly to the H band.
For two orbits, we obtained four dithered exposures each,
for a total exposure time of 5120\,sec.
The NICMOS data were reduced using the ``nicred'' package
of custom C-program and IRAF\footnote{ IRAF (Image Reduction and Analysis Facility) is distributed by
   the National Optical Astronomy Observatories, which are operated
   by the Association of Universities for Research in Astronomy, Inc.,
   under contract with the National Science Foundation.}
scripts developed for this project (McLeod 1997, Leh\'ar~et al. 1998).
We also re-analyzed two orbits of HST WFPC2 observations of \MG 
which had been acquired earlier (Keeton et al. 1998).
The target was observed for 8100\,sec through the F675W filter (R-band),
and 10500\,s through the F814W filter (I-band).
In each case, the orbit had been split into several sub-exposures
to remove cosmic ray events.  Details of our WFPC2 reduction are also
given elsewhere (Leh\'ar~et al. 1998).  The 8~GHz radio image is from
Chen \& Hewitt (1993).

Figures 1 and 2 show the PC field (R and I-band co-added)
and NIC2 field around the lens galaxy. We used SExtractor 
(Bertin \& Arnouts 1997) to catalog the objects in the fields, 
using the default settings for most software options.  The total object 
magnitudes were estimated using Kron-type automatic apertures,
and their colors were 
determined using fixed circular apertures (of diameter $0\parcs3$, $0\parcs56$, $1\parcs0$, 
or $1\parcs7$, chosen to exceed twice the rms object size along its major axis
in the I-band exposure).  In computing the object magnitudes, we used zero-points
of $22.08$ for R, $21.69$ for I (Holtzman et al. 1995, gain of $7$, corrected to
infinite aperture) and $21.80$ for H, which correspond to 1 count/sec.
We only cataloged objects in the
PC field which were detected 
in both WFPC2 bands. These objects are shown in Figures 3 and 4 and listed
in Table 1, where we have labeled them
in order of decreasing I-band flux. In our nomenclature the
main lens galaxy is G and the remaining galaxies in the PC field are
labeled G1, G2 $\cdots$ in order of increasing I magnitude.  Thus the
two nearby perturbing galaxies found by Annis (1992) are C=G15 and D=G9. 

\subsection {Photometric Models of Galaxy Evolution }

To interpret the colors and magnitudes of the galaxies detected in
the field, we used spectrophotometric models (Bruzual \& Charlot 1993)
to compute K and evolutionary
corrections for present-day $L_*$ galaxies ({\it i.e.} galaxies
of different morphological types 
with $M_B=-19.9 + 5\log h_{100}$ at $z=0$). These  corrections allow us to 
estimate the magnitudes and colors of such galaxies
at any other redshifts. For all galaxy types we assumed star-formation histories (SFH) that start at
$z_f = 5$, in three different cosmologies:
$\Omega_0=1$ ($\Lambda_0=0$), $\Omega_0=0.3$ ($\Lambda_0=0$) and $\Omega_0=0.3$ ($\Lambda_0=0.7$).
For early-type galaxies, we modeled the SFH
either by an initial ``burst'' of constant star formation
that is truncated after 1~Gyr, or by an ``E/S0'' model, which has an
exponentially decaying star formation rate, $e^{-t/1{\rm Gyr}}$.
For spiral galaxies, we used
a star formation rate proportional to the gas fraction, where
the proportionality constant was taken from Guiderdoni \&
Rocca-Volmerange (1988) and decreases from Sa to Sb to Sc.
We assumed a Salpeter IMF for the burst and E/S0 models,
and a Scalo IMF for the spiral models (see Keeton et al. 1998 for details). 
To predict the broad-band fluxes,
the resulting energy distributions were convolved with the 
F675W, F814W, and F160W filter transmission curves.

Figures 3 and 4 show the observed color-magnitude and color-color
diagrams for the cataloged galaxies, along with these
spectrophotometric models for $\Omega_0=0.3$ ($\Lambda_0=0.7$).
To make the comparison, we de-reddened the observed photometry,
accounting for galactic
foreground extinction of
$E(B-V) =0.036$ (Schlegel, Finkbeiner \& Davis 1998).  
Figure 3 also shows the
R and I photometry for galaxies detected over the full area
of the WF/PC images using open triangles, highlighting
the objects in the NIC2 field by solid points.
The curves show the predicted magnitude and color of an $L_*$ galaxy
as a function of redshift (with the line pattern changing every
$\Delta z = 0.5$).  For brighter (fainter) galaxies the corresponding curves
shift horizontally to the left (right).

\subsection{Wavelength Dependent Morphology of the System Lens}

The wavelength dependent morphology of the
\MG lens system is illustrated in Figure 5, which shows the 8~GHz map (Chen \& Hewitt 1993)
along with H, I, and R images, centered on the lens galaxy.
Note that the R and I images are slightly smoothed 
to enhance the visibility of faint features and that the images
were registered using the center of the lens galaxy ($=D_R$ at
8~GHz).
The flux fraction arising from the 
lens galaxy, the source galaxy, and the active source nucleus 
all change dramatically as a function of wavelength, even
among the optical and near-IR images.
We need to describe the morphology in detail, 
because of the complexity of structures and because 
the nomenclatures for the ring features 
differ between the optical and the radio. The radio cores
seen at low frequency (8~GHz and below) are $A_R$ and $B_R$ where 
the subscript $R$ denotes a radio feature.  At higher frequencies (15~GHz and above),
the $A_R$ core splits into two components $A_{R1}$ and $A_{R2}$, where $A_{R1}$
and $B_R$ correspond to the active nucleus (Chen \& Hewitt 1993, Chen et al. 1995).
In addition to the cores and the ring, there is an apparently
unlensed radio lobe $C_R$ to the
Southwest, and a central image $D_R$.  We see the lens galaxy G in both the optical
and infrared images, as well as the two perturbing galaxies C=G15 and D=G9.
At the locations of the radio cores we see two peaks in the infrared, 
but nothing
in the optical.  
The infrared peaks $A$ and $B$ correspond to the locations of the radio cores
$A_{R1}$ and $B$ when we register the central radio component $D_R$ on the
center of the lens galaxy.  We label the extended peaks seen in the infrared $S_A$ and $S_B$, 
and identify them with two images of the central regions of the host galaxy. 
The overall structure of the ring differs in every image.  The radio source is imaged into a 
complete, thin ring because the extended radio source covers little area
(Kochanek et al. 1989, Chen et al. 1995).  The thickness of the infrared ring means
that the extended infrared source is much larger than the radio source, and the
peak of the extended emission is clearly offset from the active nucleus. 
The optical source must be smaller and offset from both the active nucleus and the
center of the extended infrared source to produce the 
observed partial rings.

\subsection{Structure of the Dominant Lens Galaxy}

As the R, I and H images reveal, the
lens is an early-type galaxy, which is well fit by a de Vaucouleurs profile
with major axis effective radius $R_e=0\parcs68\pm0\parcs05$, axis ratio $b/a = 0.77\pm0.02$,
and major axis position angle $60^\circ \pm 2^\circ $ in all three filters.   Attempts to fit 
the profile with an exponential disk led to significantly higher residuals.  
The lens galaxy colors of I--H$=2.68$ mag and R--I$=1.28$ mag 
 (see Figures 3 and 4) are well matched to the predictions of the
initial burst or the E/S0 photometric models at a redshift of $0.8 \ltorder z_l \ltorder 1.0$,
quite close to the redshift previously estimated by Hammer et al. (1991) of $z_l=0.85$.
If we map the galaxy back onto the fundamental plane as observed in nearby clusters,
we estimate that the lens redshift is $z_l=0.89\pm0.03$ at 1--$\sigma$ ($\pm0.07$ at 2--$\sigma$).
We will adopt the Hammer et al. (1991) redshift and the initial burst photometric model
for the remainder of the discussion.
The initial burst models predict $m_H=18.0$ ($18.4$) for an $L_*$ galaxy
and $\Omega_0=1$ ($0.3$ flat), close to the observed the $m_H=18.57$
for the lens.
The effective radius is $6.4 h_{65}^{-1}$ ($8.2h_{65}^{-1}$) kpc for $\Omega_0=1$
($0.3$ flat), compared to $R_{e*} = (6.2\pm1.5)h_{65}^{-1}$ kpc scale length for an $L_*$
early-type galaxy in local samples (\eg Jorgensen et al. 1992) for $H_0 = 65 h_{65} \kms $ Mpc$^{-1}$. 

The red color of the lens best matches both the E/S0 and initial burst spectrophotometric models.
Dust in the lens galaxy probably cannot explain its colors.  For a rest frame $R_V=3.1$ 
extinction curve, the lab-frame extinction coefficients for our filters are $R_{F675W}=4.82$, 
$R_{F814W}=4.11$, and $R_{F160W}=1.59$ using the Cardelli et al. (1989) extinction curve model.  
If we uniformly mix dust into the galaxy using the simple model that the overall flux is 
attenuated by $(1-\hbox{e}^{-\tau})/\tau$ where $\tau$ is the total optical depth of the dust 
through the galaxy, we can compute the intrinsic colors and magnitudes of the lens galaxy.  
In the color-color diagram the lens shifts to the colors of a later type galaxy at essentially 
the same redshift, but in the color-magnitude diagram the galaxy becomes far too luminous for 
any galaxy at that redshift (see Figures 3 and 4).  Compared to a dust screen, dust mixed with stars has little 
effect on the colors compared to the luminosity, and more sophisticated models (e.g. Witt et al. 1992) 
for the effects of extinction tend to further reduce the color changes.   
Because the lens galaxy is already an $L_*$ galaxy, the extinction required by the ``dusty lens''
hypothesis for the red colors of the system would make the lens galaxy anomalously luminous for
any realistic extinction model. Therefore we can rule out a very dusty lens galaxy 
independently of the properties of the lensed source.

\subsection{Lensing Galaxy Group}

The galaxies G, G3, G5, G7, G8, G9=D and G15=C form an isolated
cluster in both the color-magnitude and color-color diagrams for the field
(see Table 1 and Figures 3 \& 4).  This result strongly suggests that they are
early-type galaxies, with a range of 10 in H luminosity, that form a 
group or cluster at the lens redshift.  All these galaxies have the colors of 
passively-evolving models and there are no candidates for late-type group members 
in the NIC2 field. This is consistent with the findings in rich clusters at comparable 
redshifts (Stanford et al. 1998).  In the flanking WF/PC fields we cannot distinguish 
late-type group members from lower redshift and luminosity galaxies based on the I--R 
color alone.  No similar concentration of galaxies is seen in the surrounding WF fields.  
Nearly 50\% (6 of 13) of all galaxies in the PC field with $ 20.0 < m_I < 23.5$ fall in 
the range $1.0 < \hbox{R--I} < 1.5$, while in the adjacent fields that fraction is
only 12\%, or 12 of 104 (see Figure 3).  Galaxy C=G15 is peculiar because 
of its relatively bluer R--I color and redder I--H color, and it may be a 
representative of the ``faint, red outlier galaxy'' population (Moustakas et al. 1997).   
The reddest galaxy in the field besides the lensed source is G31, with R--H$=5.3$ 
(R--K' $>$ 4.7 in Annis (1992)).

Chen et al. (1995) noted that the two galaxies C and D had the locations
and relative fluxes needed to produce the external shear required to fit the
8~GHz radio ring. However, they could not produce the required {\it amplitude} 
of the tidal perturbation if they scaled the masses of galaxies C and D to the 
main lens G using the Faber-Jackson (1976) relation and the published K' 
magnitudes (Annis 1992).  We now know, however, that the earlier IR magnitudes
for G were gross overestimates because they were ignorant of the existence of the
ring (see Figure 5), and that C and D are 26\% and 12\%
of the luminosity of G rather than 3\% and 2\% respectively.  The Faber-Jackson
(1976) relation now predicts that C and D have the masses required to produce
the amplitude and orientation of the external shear in the 
Chen et al. (1995) lens models.   

\section{Models}

\subsection{Lensing Mass Distribution}

In principle one could use the radio, near-IR and optical images of the \MG system 
simultaneously to constrain the lensing mass distribution. However, the use of the 
R, I and H images for this purpose is complicated, because the source and lens emission 
overlap in the images. Therefore, we determine the lens
mass distribution only from the lensed radio image, following 
Chen et al. (1995). However, our new NICMOS data serve as improved constraints on the positions 
and relative luminosities of the three relevant lens galaxies.  The mass model used to 
fit to the 8~GHz data consists of three singular isothermal spheres, located at the 
positions of galaxies G, C, and D. We then fit the radio ring using 6 Gaussian source
components as an approximation to a true non-parametric source reconstruction. To
keep the procedure computationally feasible we restricted ourselves to circularly symmetric 
potentials for each galaxy. The best fit yields critical radii of $b_G = 0\parcs819$, 
$b_C = 0\parcs259$, and $b_D = 0\parcs367$ for the three galaxies and the residuals 
are comparable to those of the best models found by Chen et al. (1995). 

We can estimate the tidal perturbation or external shear produced by the other
galaxies in the PC field at the location of the primary lens G by assuming that the
galaxies are at the same redshift as G and scaling their critical radii relative
to the critical radius of G (see \S3.1) with luminosity $L$ using the
Faber-Jackson (1976) relation ($b \propto L^{1/2}$).  Thus a galaxy of luminosity
$L$ at angular distance $r$ from G produces an estimated tidal shear of
$\gamma_T = (1/2)(b_G/r) (L/L_G)^{1/2}$ assuming a singular isothermal sphere
mass distribution.  The shear estimate will be overestimated if the galaxy
halo is truncated on scales smaller than the distance $r$, or if the galaxy is
a later type than the lens galaxy.  Redshift differences can lead to both
over and underestimates.  Table 1 gives the tidal shears for
the objects in the PC field using the I magnitudes to estimate the relative
luminosities.  The $\gamma_T$ values give a simple synopsis of which galaxies
may be important as perturbations, although not all galaxies with large values for
$\gamma_T$ will be real sources of perturbations.  For example, galaxy G1
is an early-type galaxy at a significantly lower redshift whose perturbation
strength is greatly exaggerated.

For the group members the shear estimates should be reasonably accurate.  Equivalently
we can predict the relative velocity dispersions (or critical radii) of G, C, and D 
from their relative luminosities, and with the newly determined magnitude of G we would expect
to find  $b_C=0\parcs35\pm0\parcs08$ and $b_D=0\parcs26\pm0\parcs06$ for $b_G = 0\parcs82$.
The uncertainties are based on the observed scatter of 0.5~mag in the luminosity-separation 
relation (Keeton et al. 1988), and we have used the H band flux ratios rather than the
I band flux ratios.  Thus the luminosity-predicted critical radii
are in reasonable agreement with the ones derived directly from 
the lens model.  The colors of galaxy C are peculiar (see \S2.5) so it is not
surprising that it shows a larger discrepancy than galaxy D.  We can
quantify the tidal fields of the galaxies by their external shears of
$\gamma_C=0.049$ and $\gamma_D=0.054$ or their total combined shear of
$\gamma_{CD}=0.077$. The remaining group galaxies individually contribute 
$\gamma_T=0.01$ to $\gamma_T=0.02$ but their tensor shear contributions largely cancel, 
leaving a net additional shear of only $\gamma_T < 0.01$.  Moreover, they are 
$50 h_{65}^{-1}$ to $75 h_{65}^{-1}$ kpc distant from the lens, so the likely tidal 
truncation of the galaxy halos on the scale of the member separations would further 
reduce their contribution.  

In contrast to Chen et al. (1995) we represented the galaxies C and D by singular
isothermal spheres rather than the lowest order term in a tidal expansion of the
gravitational potential.  The higher order terms
significantly distort the critical lines and caustics of the lens (see Figures 6 and 7).
They do not, however, solve the problem that the model for the radio ring is
slightly rounder than observed.  Presumably, the solution lies in using models 
with both the external galaxies and an ellipsoidal model of the lens galaxy,
since Keeton et al. (1997) found that such models generically produced
good fits to the four-image lens systems. Such lens models for extended sources
are computationally expensive and are beyond the scope of this paper.
However, we deem our models adequate for the current purpose of 
understanding the basic structure of the \MG lens system.

\subsection{Reconstructing the Extended Source Components}

We now fit the H, I, and R images using a de Vaucouleurs model of fixed structure for 
the {\it light} from the lens galaxy and the lensing mass distribution from \S3.1 to
produce lensed images of the models for the extended emission of the host galaxy.  The
H band ring is well modeled as the image of an exponential disk offset from 
the radio core.  The intrinsic magnitude of the exponential disk is H$=19.6$ mag 
with a scale length of $R_d=0\parcs29$, axis ratio $0.69$ and major axis PA of
$70^\circ$.  The axis ratio corresponds to an inclination angle of $44^\circ$
(where $90^\circ$ is face-on).  For comparison, the intrinsic magnitude of the
source corresponding to the radio core is H$=22.8$ mag.  The core component is
offset by $0\farcs2$ to the Northwest of the center of the exponential disk. The residuals from this
model consisted of excess emission on the Northeast side of the ring, which we
could model and subtract by the addition of several extended Gaussian components
to the source model with a total magnitude of H$=21.9$ mag.  The total magnitude
of all the source components was H$=19.3$ mag, which we see as a ring with a total
magnitude of H$=17.4$ mag.  Figure 6 shows the ring structures after subtracting
the best fitting model of the lens galaxy, and Figure 7 shows the reconstruction of 
the unlensed source.  We clearly see the offset between the peak of the 
extended stellar emission and the AGN core.  The major axis of the radio jet is 
roughly perpendicular to the offset vector.   

The I and R rings are both more complicated in structure and significantly
noisier, so it is difficult to evaluate which small scale substructures of the optical
ring are real.  We decided to use a smoothed version of the image (as shown
in Figures 5 and 6) to enhance the low surface brightness images of the host
galaxy.  The key geometric fact is that the extended arc seen in the I image 
(Figure 6) running from South of the lens to the Northeast predicts the compact 
piece of the ring seen to the West of the lens.  Similarly, the gaps in the ring 
near the locations of both radio cores correspond to lensed images of the same source 
regions.  No counterparts of the radio cores are visible in the optical images.  
{\it The structural features of the optical ring are features intrinsic
to the host galaxy rather than features produced by absorption in the lens galaxy.}  

Since the optical ring corresponds to only portions of the infrared ring, we
modeled the source by a grid of $\sim 300$ extended Gaussians whose parameters 
were optimized to match the observations.  
The model correctly reproduces the
ring structure but is limited by the poor signal-to-noise ratio between the 
ring and the background.  Figure 7 shows the reconstructions of the source 
morphology in I and R.  The intrinsic magnitudes of the host are I$=23.2$ mag
and R$=23.8$ mag.   
The R and I-band flux comes mostly from the SE half of the host galaxy, with a 
sharp drop in the emission starting near the peak of the extended IR emission 
and extending over the location of the AGN core.  The changing morphology of 
the lens between the optical and the infrared appears to be entirely created 
by changes in the morphology of the source.  Much of the optical emission lies 
along the radio jet axis, which is suggestive of the alignment effect (McCarthy 
et al. 1987). Assuming $z_s > 2$, the aligned emission represents UV rest-frame
light. If \MG is typical of other high redshift, high luminosity radio sources,
the alignment is caused by a combination of off-axis scattering of the core light
and jet-induced shocks in the ISM of the host galaxy (Tadhunter et al. 1992, 
van Breughel et al. 1985).

Alternatively, the geometry is suggestive of an AGN embedded in a larger, dusty 
stellar disk (e.g. Centaurus A).  The infrared source is relatively well modeled
as an exponential disk at an inclination near $45^\circ$ with the far side of the disk 
in the Southeast quadrant.  Qualitatively, at least, such a geometry could both hide half 
the galaxy at the shorter optical wavelengths and produce an offset
between the radio core and the apparent center of the host galaxy.  The peak of the
extended emission does appear to steadily shift to the Southeast for the bluer
wavelengths.  The reconstructed morphology of the optical source does not perfectly
correspond to such a model, but the data used for the reconstructions are also poor.
The complicated morphology, the quality of the optical data, and the lack of a source
redshift make it impossible to interpret the colors of the source in detail.  The
average color is ``bluish'' in the optical, with R--I $\simeq 0.6 \pm 0.3$ mag, and 
red from the optical to the infrared, with R--H $\simeq 3.9 \pm 0.3$.  
Larkin et al. (1994) found that the brightness of the source
increased dramatically between J and H, but modestly between H and K which
suggests that the source redshift is above 1.8 and below 3.0 if we assume
that the spectrum has a break near 4000 \AA .  The mean colors of the
source are roughly consistent with a stellar population at $z_s > 2$ (see Figure 4), 
but they impossible to interpret in detail without a good model for the effects of 
the dust in the source on the fluxes.  We obtain a similar estimate of $z_s \gtorder 2$ 
if we require the lens galaxy to lie on the fundamental plane (either fixing the lens 
redshift to the Hammer et al. (1985) value or performing a simultaneous estimate).

\section{Consequences of the Limits on the Dust Content of the Lens}

The residuals in modeling the H-band ring provide the best means of limiting the amount
of differential extinction produced by the lens galaxy.  The surface brightness
of the source is unaffected by lensing, so any residuals in the H image
are a combination of noise, shortcomings in either the photometric or 
lens model, and differential extinction.  The fractional
residuals $f$ in the H image are related to the differential
extinction by $\Delta E(B-V) = (5 f/R_{F160W} \ln 10)(1+f^3/3+\cdots)\simeq 1.2 f$ 
over the multiply-imaged region.  Figure 8 shows the fractional residuals
(modestly smoothed) for the regions exceeding 10\% of the peak flux in the
ring.  Our model makes the smallest errors in the bright parts of the
ring, where $|f| \ltorder 5\%$, and then rises near the edges where
we reach the limits of our parameterization of the surface brightness
distribution.  Thus over the region from $0\parcs5 \ltorder R \ltorder 1\parcs5$ 
from the lens center ($R_e \ltorder R \ltorder 3R_e$ or 
$3 h_{65}^{-1}\hbox{kpc} \ltorder R \ltorder 10 h_{65}^{-1}\hbox{kpc}$)    
the extinction varies by less than $\Delta E(B-V) \ltorder 0.06$. 
For the patchy extinction seen in most galaxies, such a limit on the
differential extinction essentially corresponds to a comparable limit
on the mean extinction. Note that this test is also sensitive to patchy, but
grey extinction. A smoothly distributed dust component
with a significant optical depth, such as $E(B-V) = E_0 e^{-R/R_0}$, 
must have an enormous scale length to keep the ring colors as uniform 
as observed.  The limit roughly corresponds
to $R_0 > 100 E_0 h_{65}^{-1}$ kpc, which is implausible for high 
optical depths given a stellar scale length of only $6h_{65}^{-1}$ kpc.
Figures 3 and 4 also show the source reddening vector
given up to $E(B-V)=1$ mag of extinction in the lens galaxy.  Given the observed
source flux, the implied intrinsic source luminosity becomes implausibly large if the
lens contains any significant amount of extinction.

The existence of the infrared rings in both \MG and B~1938+666 (King et al. 1998) means
that the Malhotra et al. (1997) and Larkin et al. (1994) hypothesis that the red color of
the lenses is created by dust in the lens galaxy is ruled out.
In fact, the Annis \& Luppino (1993) hypothesis that the red color 
is created by dust in the source galaxy obscuring the AGN combined with emission by
the stars in the host galaxy is correct for both systems. 
Only MG~0414+0534 is left as a candidate for a very dusty, 
early-type lens.  While the quasar spectra in MG~0414+0534 are very reddened 
(Lawrence et al. 1995), the differential extinction between the four lines of sight 
is less than 10\% of the total (McLeod et al. 1998a) even though the paths are separated 
by $\sim 5$ kpc.  
Such uniformity in the interstellar medium of the lens galaxy seems 
implausible given the observed properties of nearby galaxies.  In fact, the
discovery of the extended blue arc connecting the brighter three quasar images
in MG~0414+0534 (Falco et al. 1997) also rules out the dusty lens hypothesis
in this system.  The arc is already bluer than the lens galaxy and has
an estimated intrinsic luminosity of $\sim 0.5 L_*$ before any correction 
for dust in the lens galaxy.  If the extinction in the lens was 
genuinely $A_V \sim 7$ mag, then the arc color would be remarkably blue
and the total arc luminosity phenomenal.  It is important to include the large Galactic
extinction of $E(B-V)=0.30$ mag (Schlegel, Finkbeiner \& Davis 1998) towards
MG~0414+0534, because it is largely responsible for the
anomalously red color of the lens galaxy.  
  
These results do not imply, however,  that complete absence of dust in lens galaxies.  
Falco et al. (1997) demonstrated that reconciling the statistics of optical and 
radio lens samples required dust in even the early-type galaxies.  They 
estimated a mean extinction for the early-type lens galaxies of $A_B=0.5\pm0.4$ 
mag in the observers frame, and that spiral galaxy lenses may be completely 
eliminated from the quasar sample.  The cosmological limits purely from the 
radio-selected sample were $\lambda_0 < 0.65$ at 2--$\sigma$ in flat cosmological 
models, so the effects of extinction provided no escape from the lensing 
constraints on the cosmological constant.  Falco et al. (1997) also noted that the
composition of the typical optical counterparts of flat-spectrum radio sources 
changed rapidly below 250 mJy
(at 8~GHz), from over 90\% quasars at brighter fluxes to approximately 50\% at 50 mJy.  
Hence the source population from which the radio lenses are drawn is qualitatively 
different from that of bright radio sources.  Although \MG appears to be nearly
transparent, several other systems are known to contain significant quantities of 
molecular gas and produce large differential extinctions of the lensed images. 
The two lenses with the largest differential extinctions, PKS~1830--211 with
$\Delta E(B-V)\simeq2.9$ mag and B~0218+357 with $\Delta E(B-V)\simeq0.8$, are also
the two lenses in which significant column densities of molecular gas are 
detected in absorption (see Gerin et al. 1997 and references therein).  
The remaining lenses appear to have far more modest
differential extinctions (see Leh\'ar et al. 1998).

\section{SUMMARY}

By combining NICMOS and WFPC2 HST images in R, I, and H of \MG with the existing 8~GHz
radio maps, we have been able to construct a comprehensive picture of both
the source and the lensing structures in the system. Our new data show that 
a correct interpretation of the system has not been possible on the 
basis of lower resolution data. In particular, we find that
the host galaxy of the source AGN is the dominant component of the 
H-band flux.
On the basis of these new data and of extensive modeling we have been
able to establish the following results:

\noindent (1) The very red optical -- IR colors of the lensed source
are due to the extended flux from the source host galaxy at $z_{source}>2$.
They are {\it not} due to dust reddening of the AGN at either the source or the
lens redshift. 
In addition, there is direct evidence from our data that
the MG~1131+0456 lens galaxy is indeed nearly transparent, in contrast
to earlier inferences by Larkin et al. (1994) and Malhotra et al. (1997).
The second reddest lens, B~1938+666,
shows very similar properties (King et al. 1998), although the smaller
angular size and more comparable luminosities of the lens galaxy and the 
ring make it more difficult to model the system in detail.
Thus we conclude that the dusty lens hypothesis of Larkin et al. (1994)
and Malhotra et al. (1997) is invalid.

\noindent (2) The new images have allowed us to establish the correct
R, I and H magnitudes for the lens galaxy and its two close neighbors.
These data corroborate that the lens is a member of a galaxy group,
or poor cluster, at $z_l\simeq 0.85$. The measured effective radius, luminosity
and mass of the main lens galaxy make it one of the highest redshift
galaxies that can be put on the Fundamental Plane. The two neighboring
galaxies (C and D) now have just the right luminosities and positions to induce the
required lensing shear, both in direction and in amplitude.
This intermediate density lens environment is similar to that of many 
other lenses (e.g. MG~0751+2716 (Leh\'ar et al. 1997), 
Q~0957+561 (Young et al. 1981), PG~1115+080 (Impey et al. 1998), and
B~1422+231 (Yee and Ellingson, 1994)).

\noindent (3) Using the lens model derived from the radio data, we reconstruct
the source structure at R, I, and H.  The bright H band ring ($m_H\simeq 17.4$ mag!)
is almost completely modeled as the image of an inclined ($45^\circ$) exponential disk
with a point source at the location of the radio core.  The radio core is offset
by $0\farcs2$ NW of the center of the extended emission.
The reconstructed source images in R and I however, show a severe
flux deficit on the NW side of the center, which appear in the observations as gaps
in the optical rings.  We argue that this morphology 
could arise from embedding the source in a larger, thin dusty disk (e.g. Centaurus A).

Few high redshift, red galaxies have been studied in detail because of the difficulties
in studying them spectroscopically.  The complicated source morphology of \MG is not
consistent with the simple model of an old stellar population at intermediate redshift
that seems to explain LBDS~53W091.  It could be consistent with the dust obscured star
formation model that explains HR~10.  A good infrared spectrum of the source 
or high resolution images of the source at J and K to complement our H band images
will be required to understand the source in greater detail.  Although NICMOS imaging is 
preferred, new, high resolution ground-based images may suffice because the H band
morphology can be used as a template for understanding any new data.
  
\acknowledgments
Support for the CASTLES project was provided by NASA through grant numbers 
GO-7495 and GO-7887 from the Space Telescope Science Institute, which is operated
by the Association of Universities for Research in Astronomy, Inc.
CSK and CRK were also supported by the NASA Astrophysics Theory Program grant NAG5-4062.
HWR is also supported by a Fellowship from the Alfred P. Sloan Foundation.

\clearpage

\clearpage
\centerline{\psfig{figure=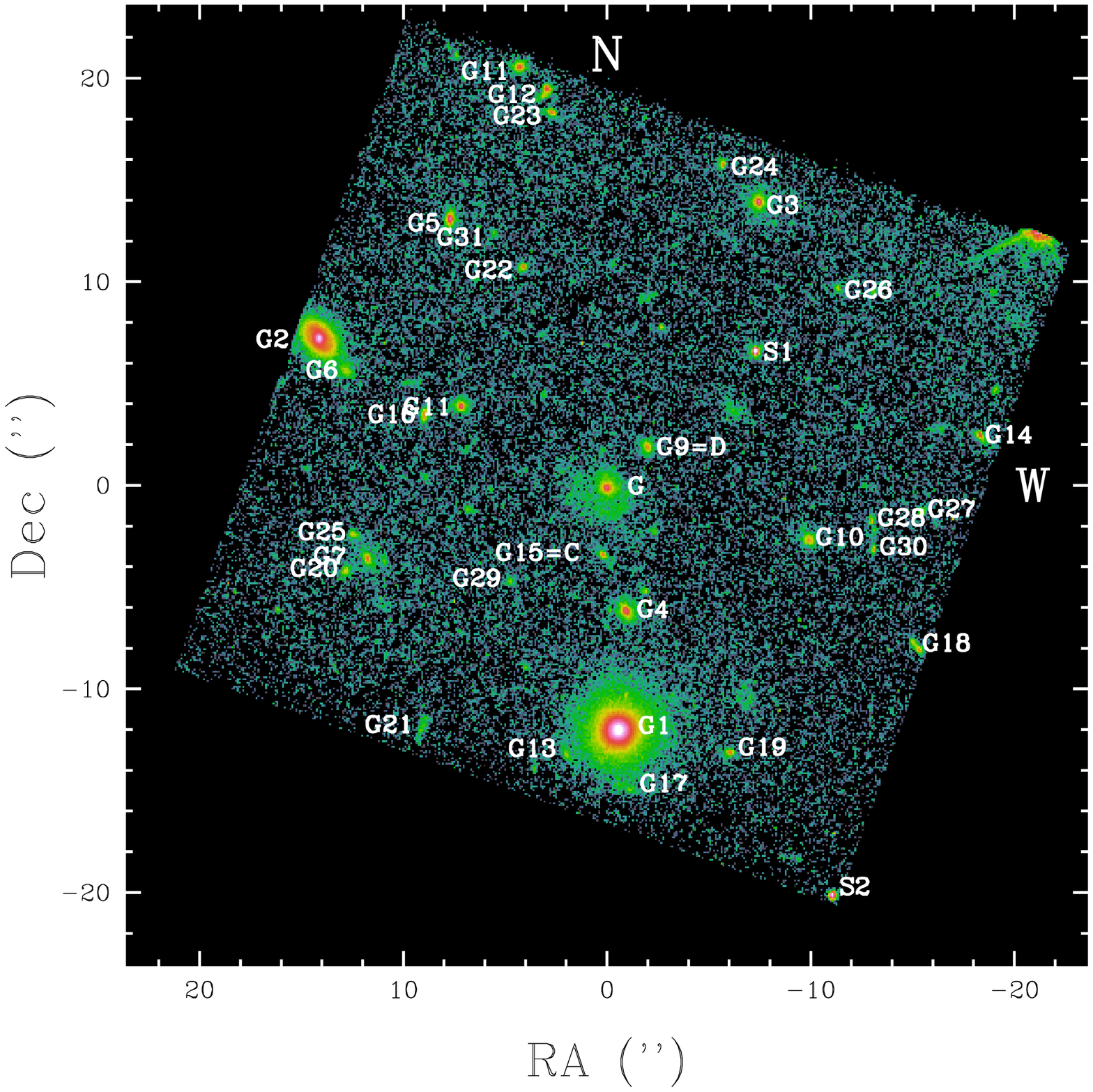,height=7.0in}}

\figcaption{Coadded I and R PC images of \MG.  Objects brighter than
  $I < 25$ mag are labeled as they appear in Table 1. }

\clearpage
\centerline{\psfig{figure=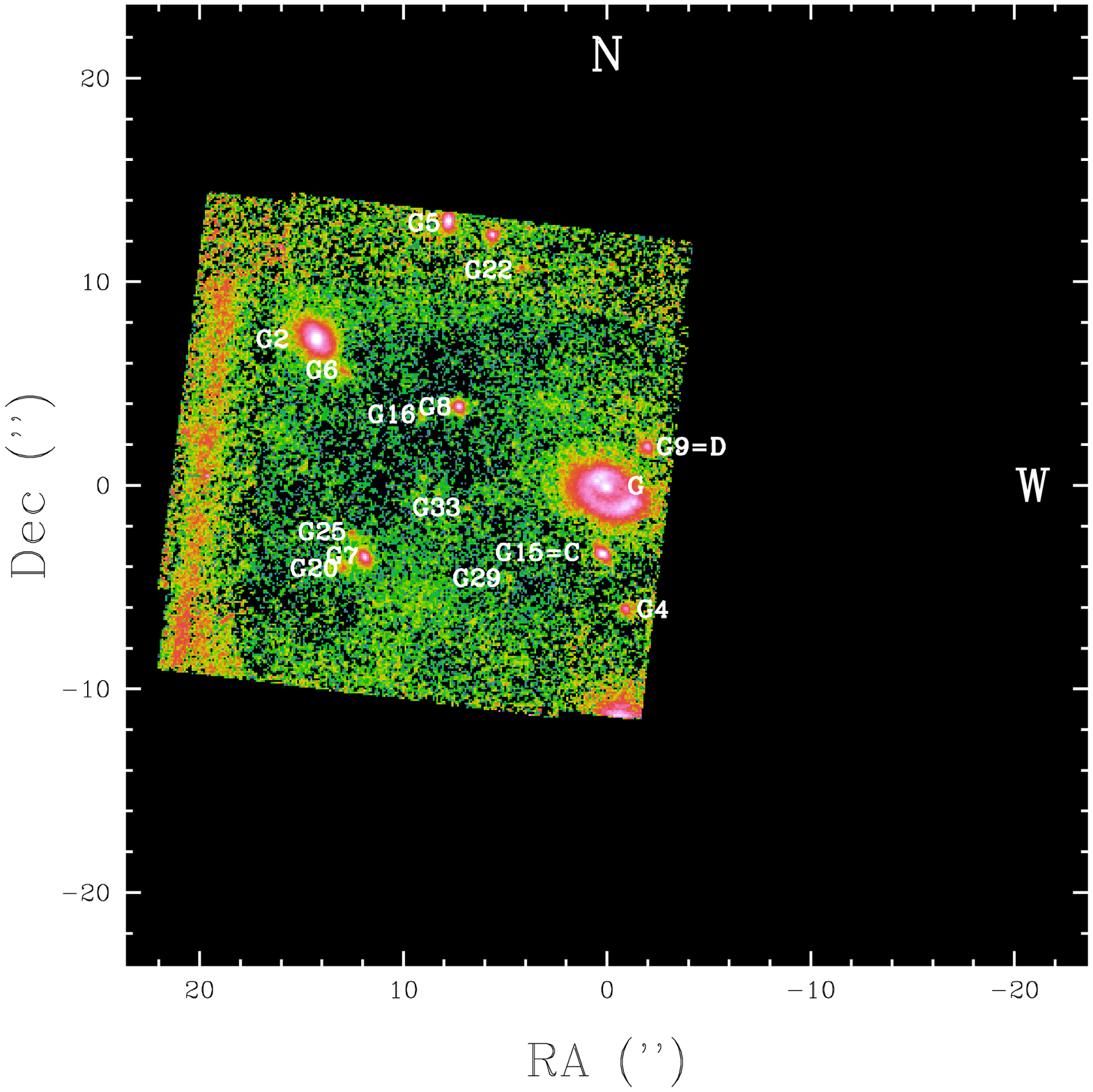,height=7.0in}}

\figcaption{NICMOS H image of \MG.  Objects brighter than $I < 25$ mag
  are labeled as they appear in Table 1. The noise increases greatly at the
  edges where the image is the sum of fewer dithered sub-images.  }

\clearpage
\centerline{\psfig{figure=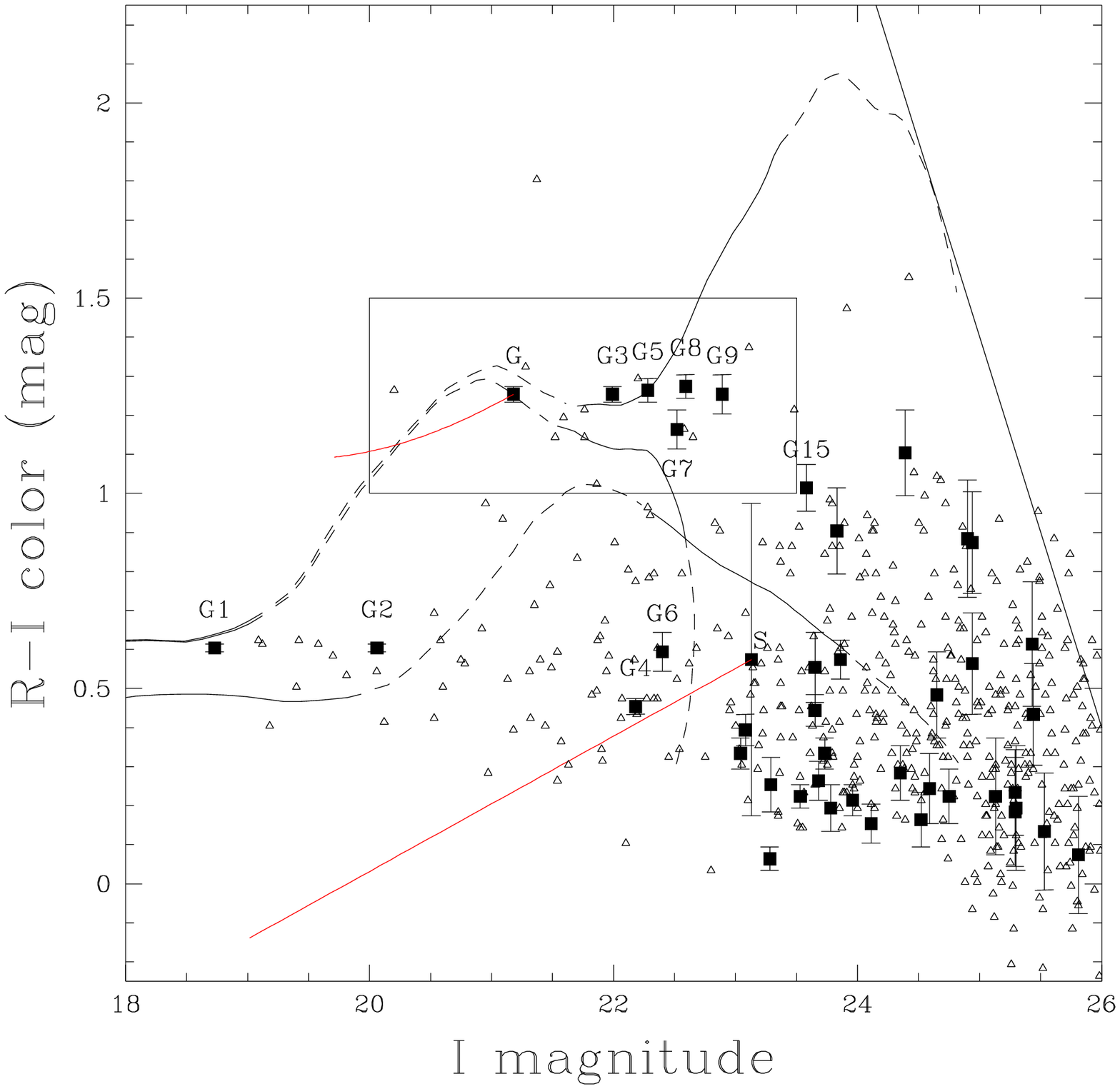,height=5.5in}}

\figcaption{R--I versus I color magnitude diagram.  The filled squares
  mark the galaxies found in the PC image, and the smaller triangles
  mark the galaxies found in the surrounding WF images.  The diagonal
  line on the right edge of the distribution shows the detection
  threshold created by the R flux limit.  The three curves show the
  magnitude and color of an $L_*$ galaxy in the burst (top), E/S0 (middle), and Sc
  (bottom) spectrophotometric models.  The line style shifts from solid to dashed
  every $\Delta z =0.5$ ($z < 0.5$ solid, $0.5 < z < 1$ dashed $\cdots$).  A
  brighter (fainter) galaxy would lie to the left (right) of the curve
  at a fixed color.   The box shows the region used to quantify the presence
  of the lensing group in \S2.5.
  The effects of dereddening uniformly mixed dust within the lens
  (up to $E(B-V)=1$ mag) is shown by the line extending from the primary
  lens G, and the effects of the same dust on the source by an analogous
  line starting at {\sl S}.  All magnitudes are corrected for the
  estimated foreground Galactic extinction. }

\clearpage
\centerline{\psfig{figure=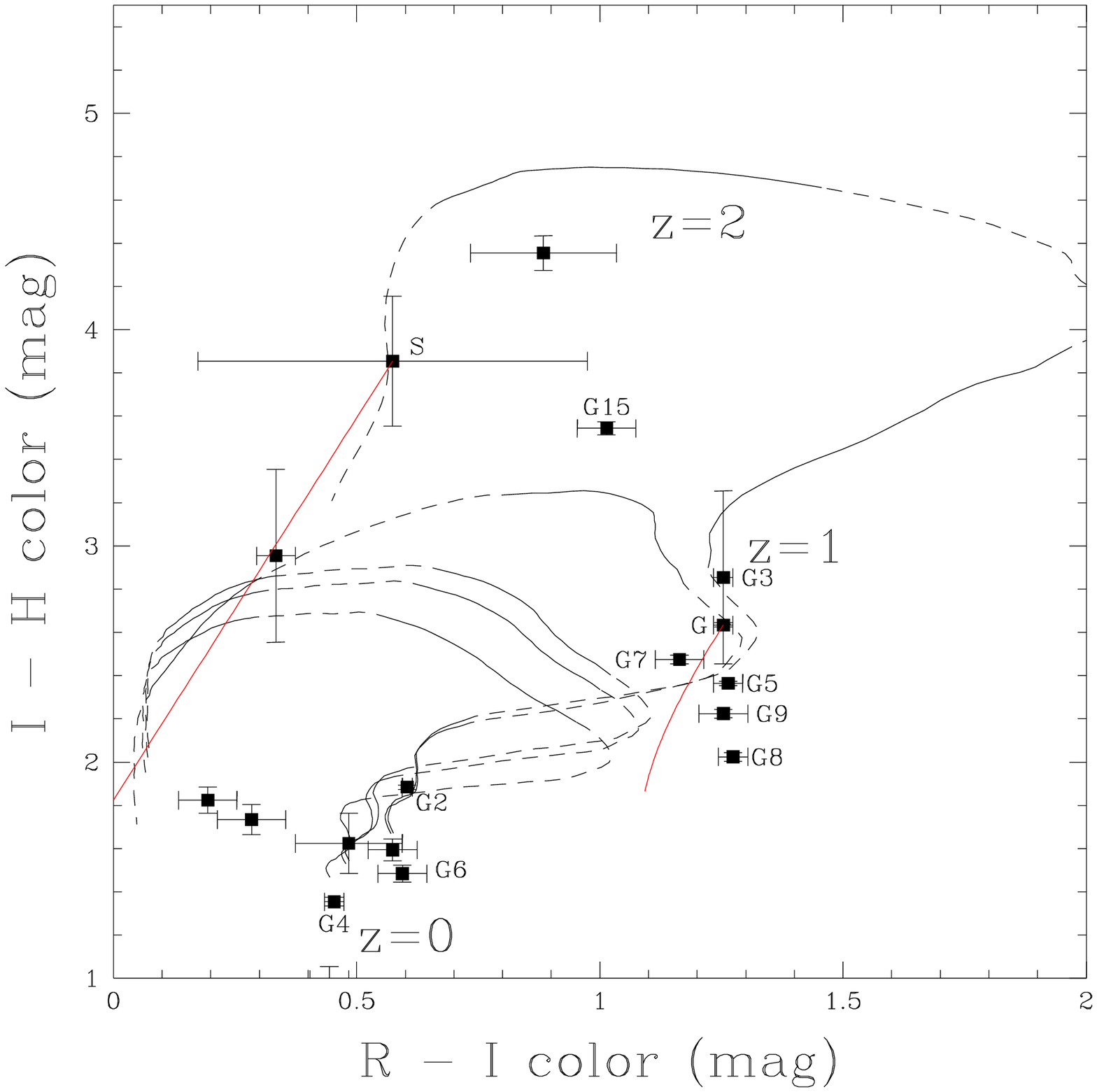,height=5.5in}}

\figcaption{I--H versus R--I color--color diagram for the galaxies found in the NIC2 image with
$I < 25$ mag.  
The five curves are the colors of the initial burst, E/S0, Sa, Sb, and Sc galaxy models. 
The line pattern shifts from solid to dashed every $\Delta z=0.5$. 
The curve extending from the point for the 
primary lens G shows the effects of  dereddening the lens galaxy by up to $E(B-V)=1$ 
magnitudes of uniformly mixed dust, and the curve extending from source (labeled S) 
shows the effect of the same dust on the source.  All magnitudes are corrected for 
the estimated foreground Galactic extinction. }

\clearpage
\centerline{\psfig{figure=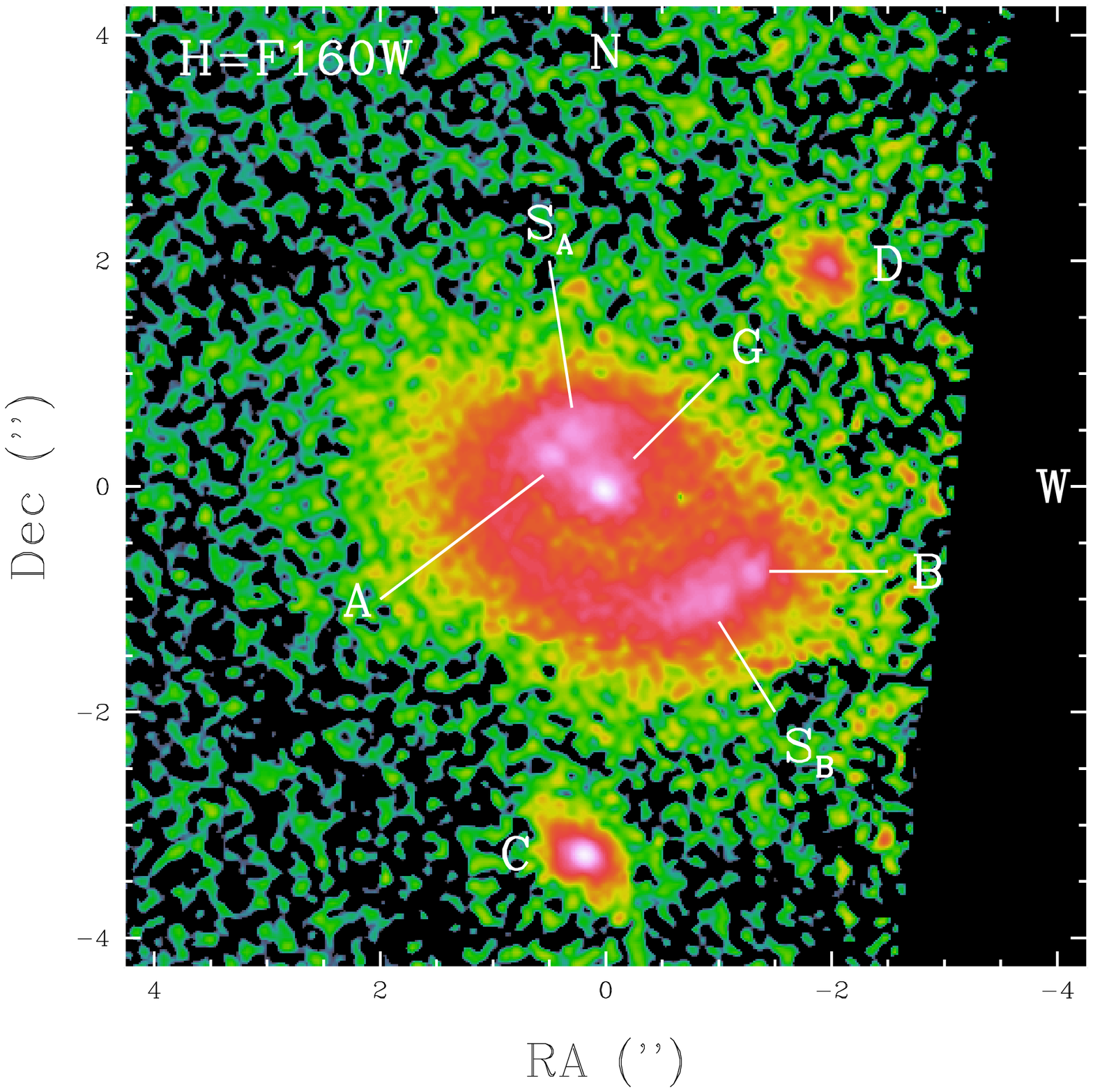,height=3.5in}\psfig{figure=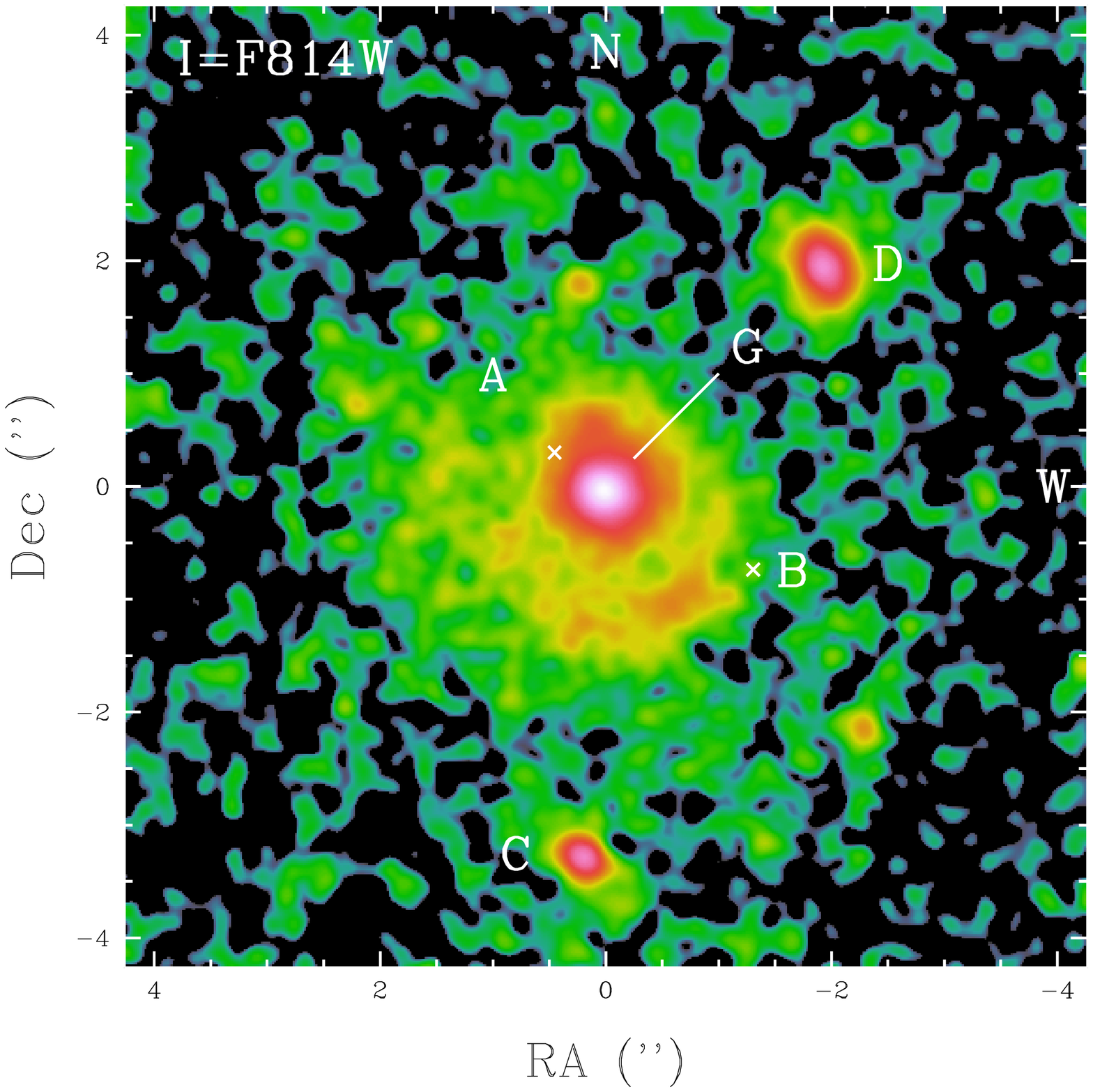,height=3.5in}}
\centerline{\psfig{figure=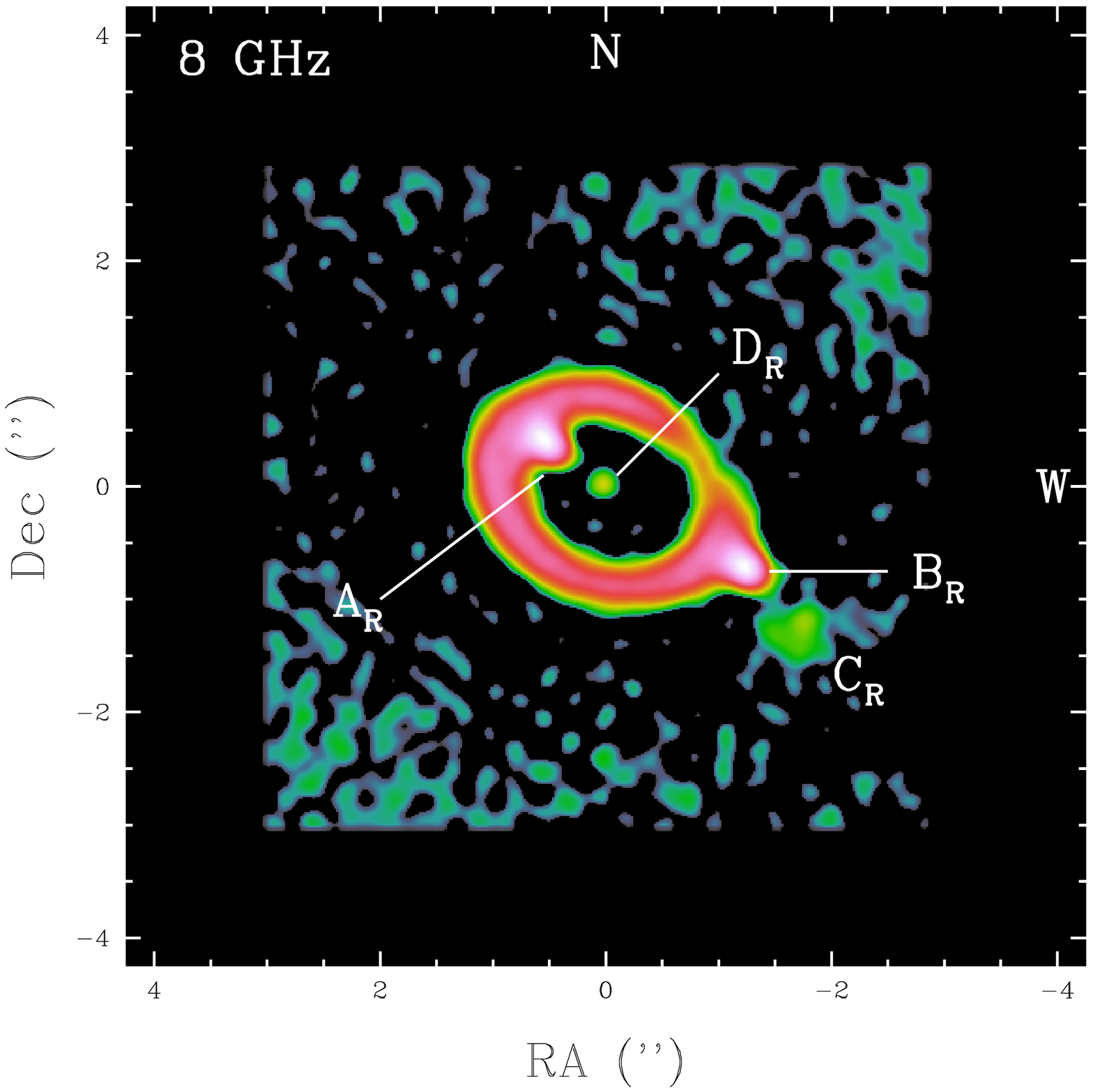,height=3.5in}\psfig{figure=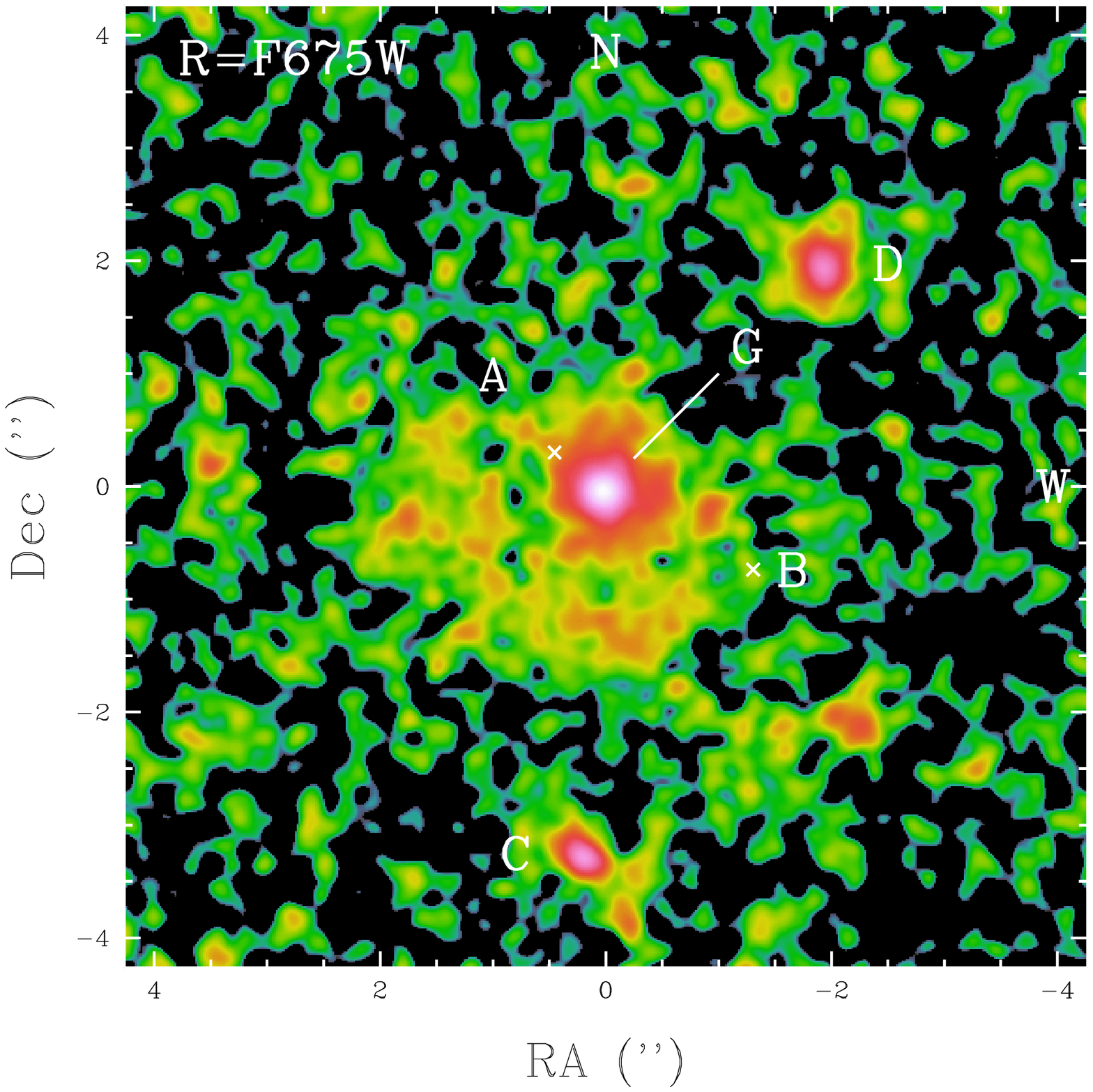,height=3.5in}}

\figcaption{Close-ups of the H (top-left), I (top-right), R (bottom-right) and 8~GHz
   radio map (bottom-left, from Chen \& Hewitt 1993) images.  The I and R images have
   been smoothed to make the ring clearly visible.  The AGN cores are located at
   A and B (A$_R$ and B$_R$ in the radio image) and they are marked by the crosses
   in the I and R images where no core component is directly visible.  The radio
   image was registered with the optical by centering component D$_R$ on the lens
   galaxy G. }

\clearpage
\centerline{\psfig{figure=fig6Hcoluse.ps,height=3.5in}\psfig{figure=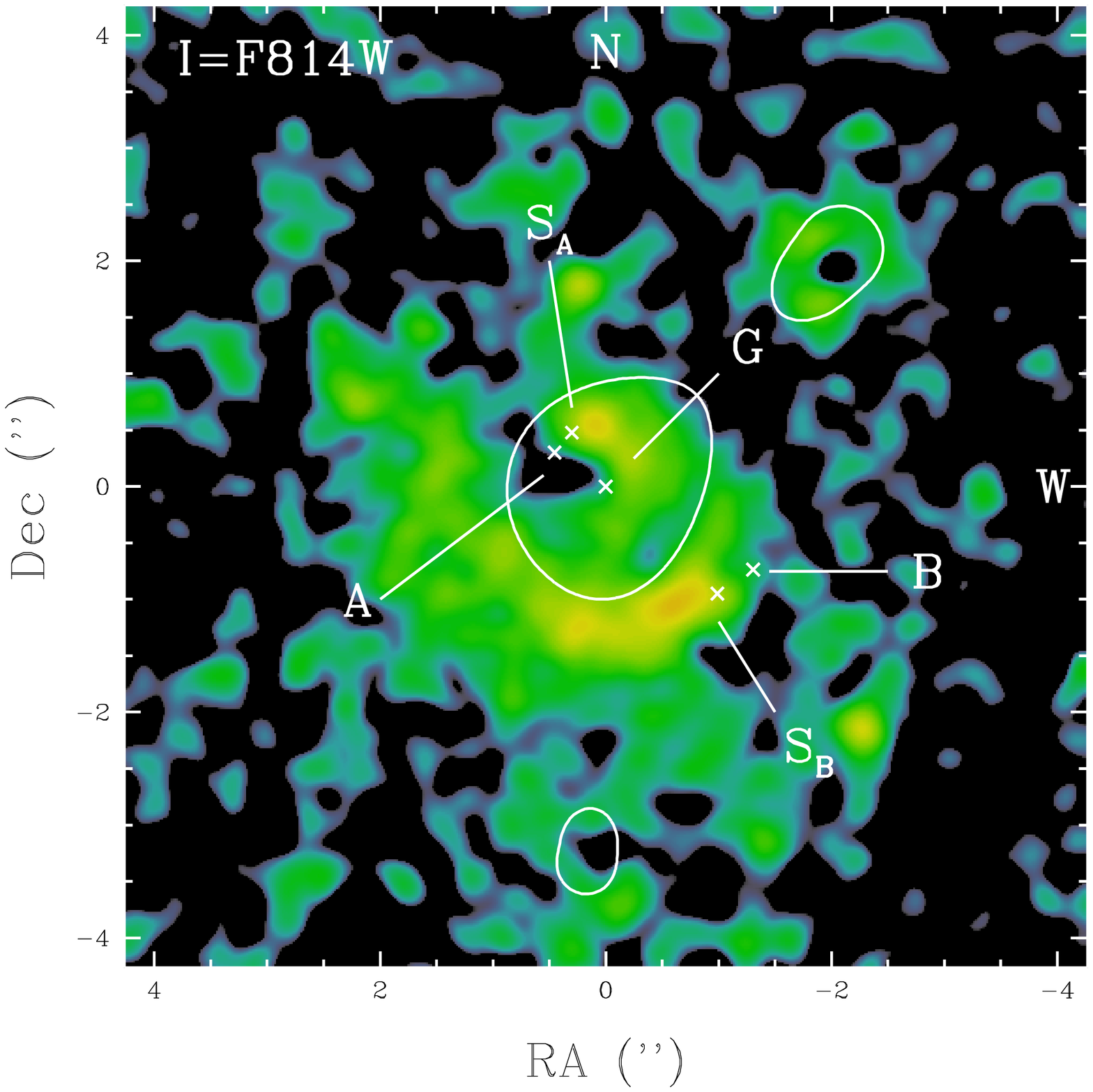,height=3.5in}}
\centerline{\psfig{figure=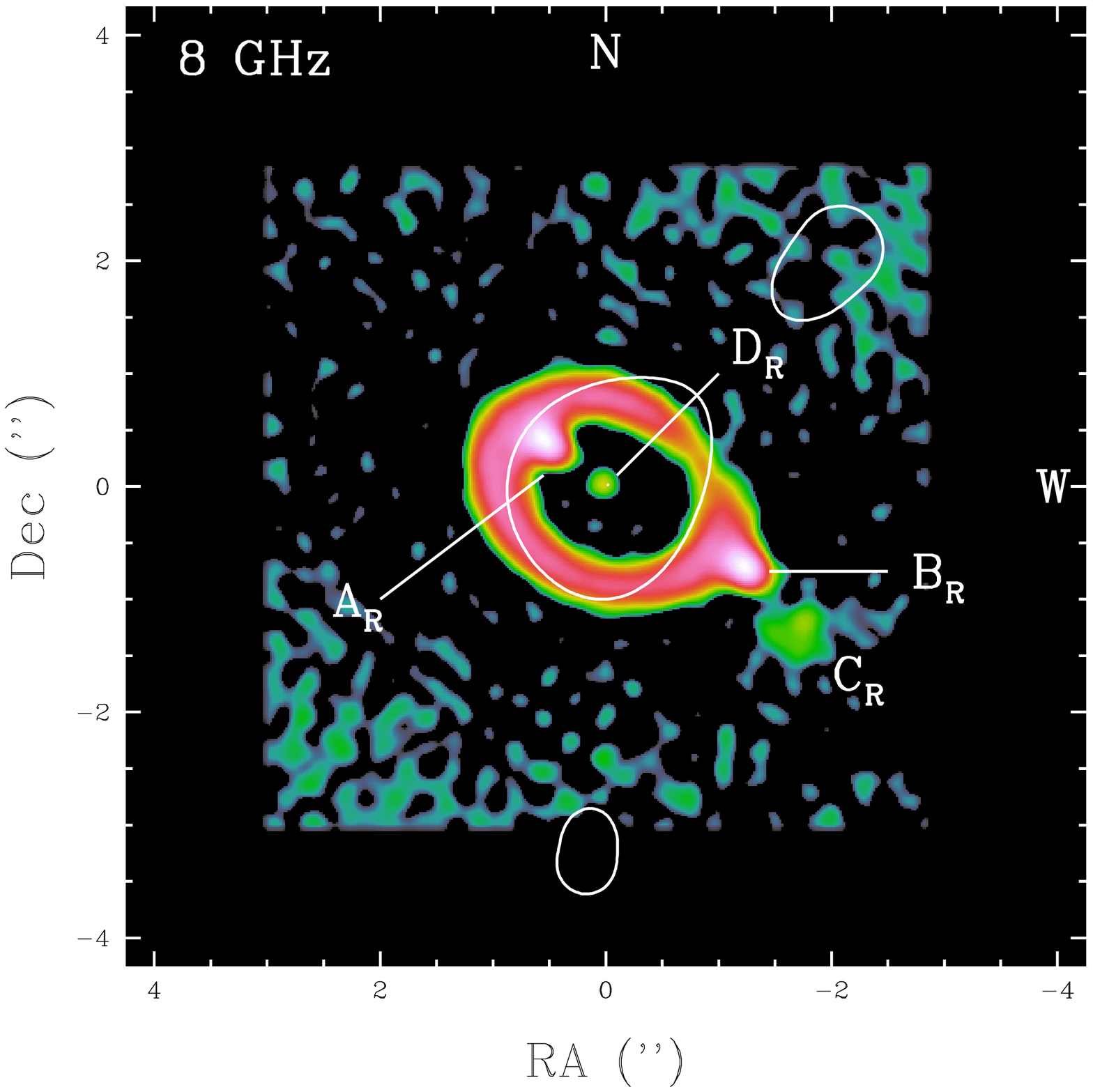,height=3.5in}\psfig{figure=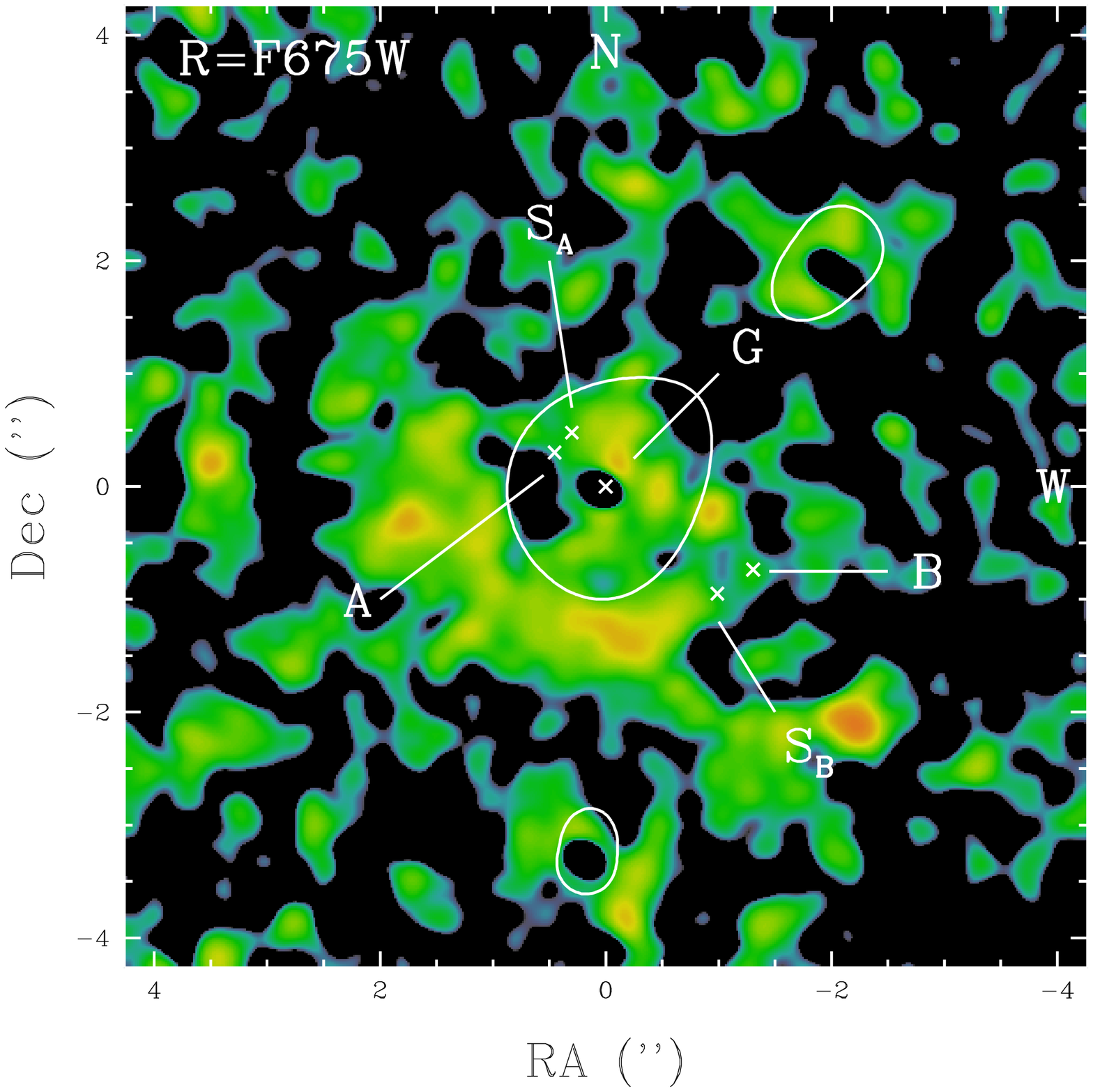,height=3.5in}}

\figcaption{Close-ups of the ring at H (top-left), I (top-right), R (bottom-right) and 8~GHz
   radio map (bottom-left, from Chen \& Hewitt 1993) after subtracting the best fit galaxy
   model.  The I and R images have been smoothed to make the ring clearly visible.  The
   contours show the critical lines of the lens model. }

\clearpage
\centerline{\psfig{figure=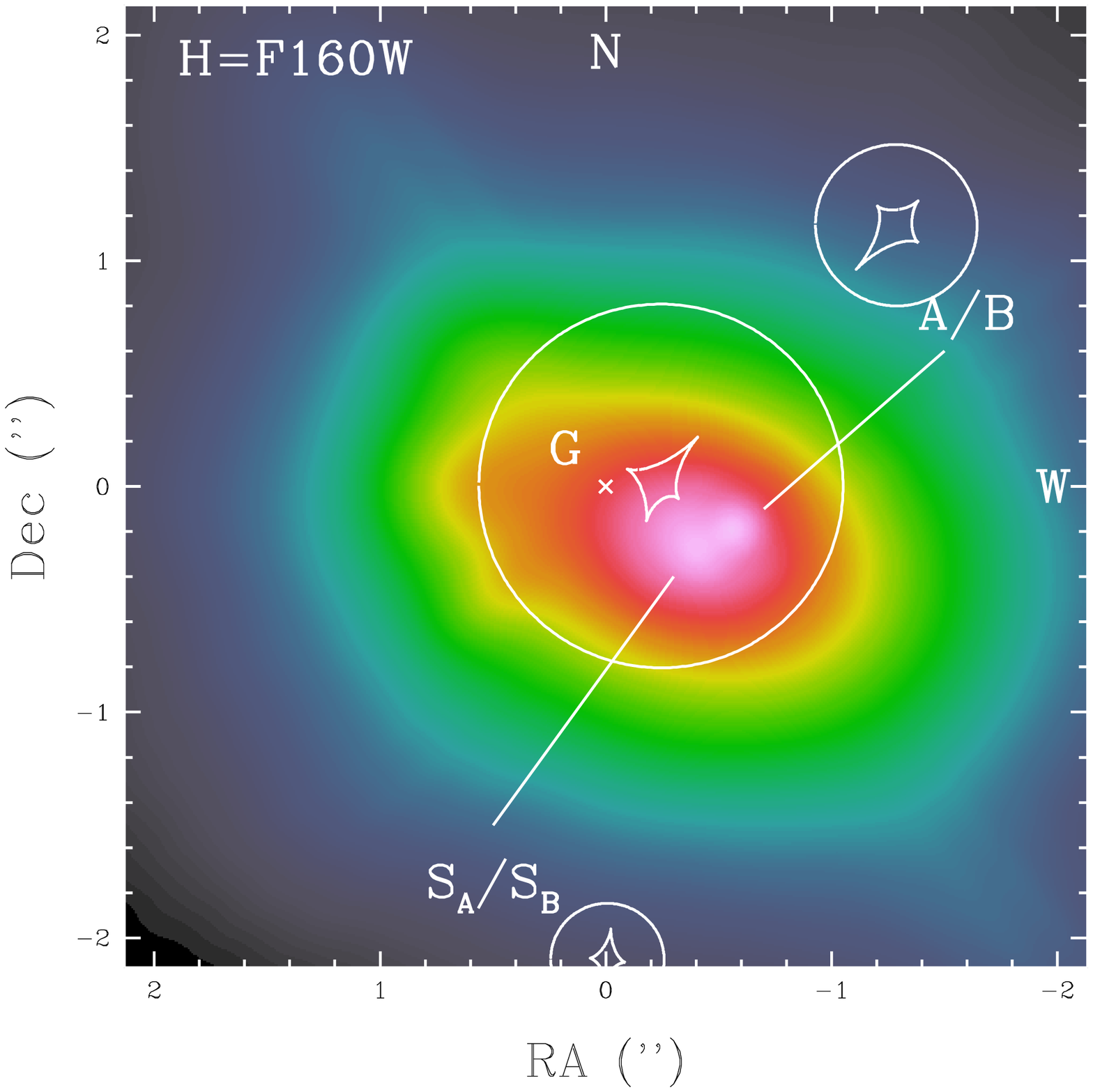,height=3.5in}\psfig{figure=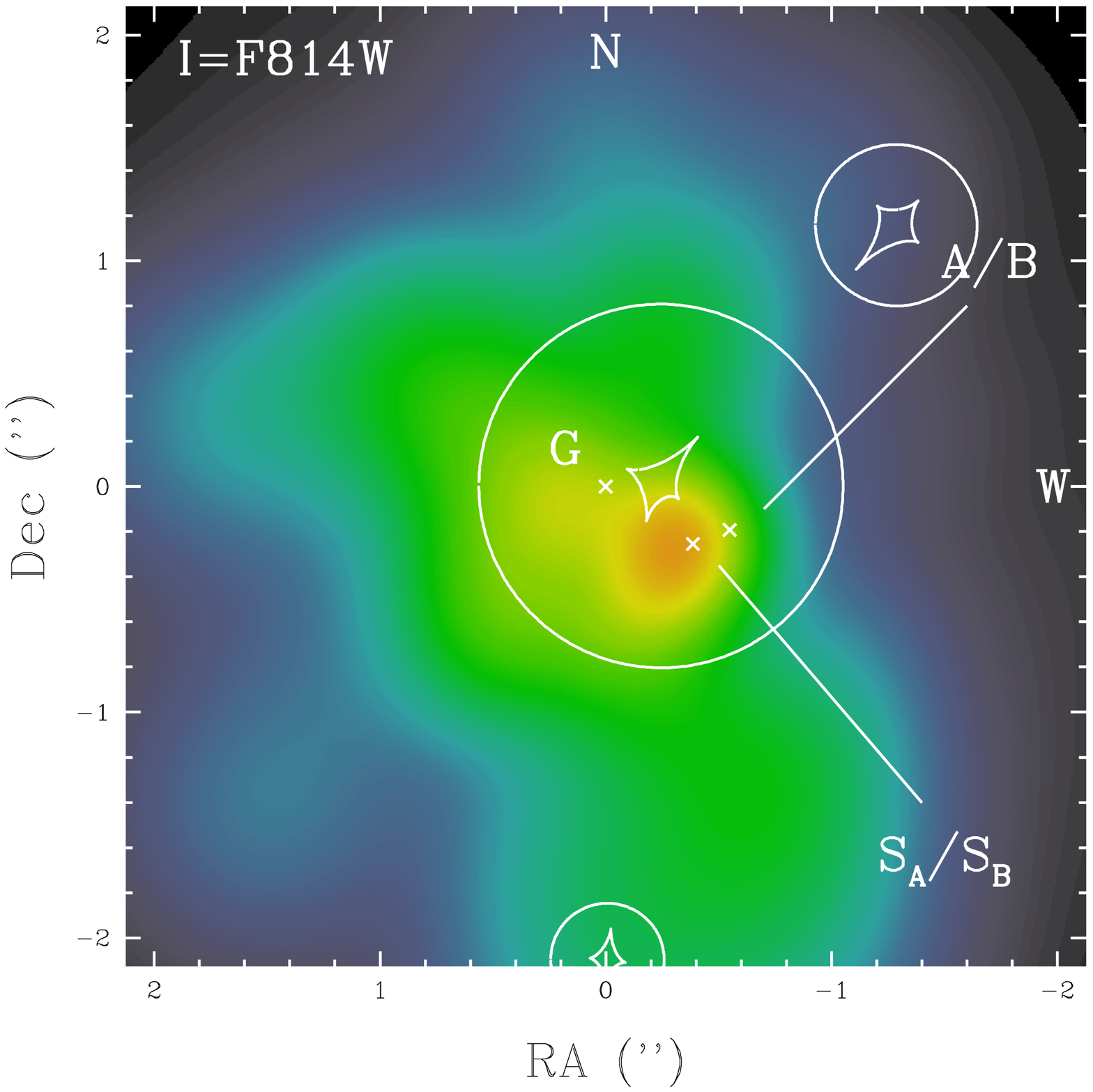,height=3.5in}}
\centerline{\psfig{figure=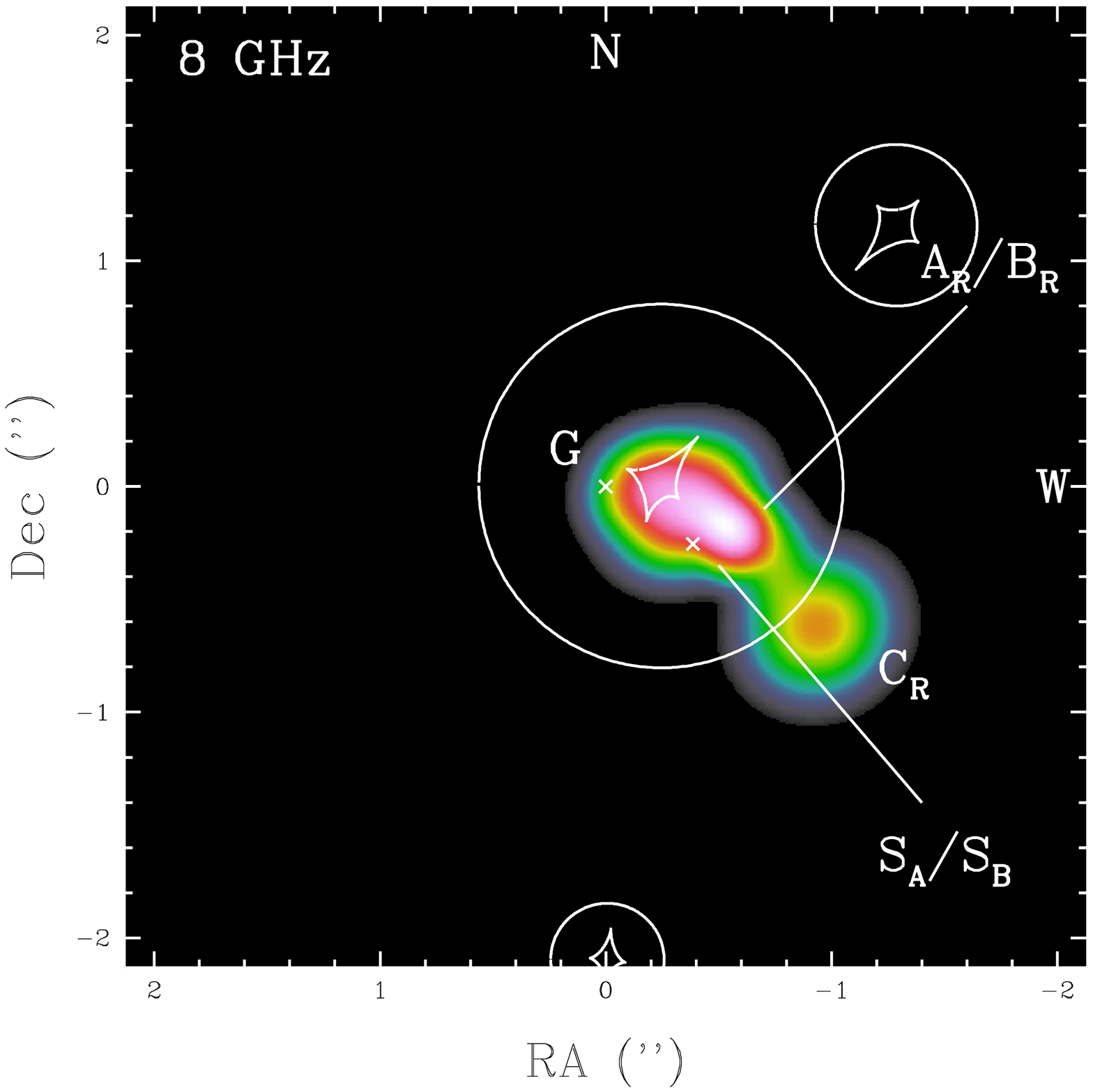,height=3.5in}\psfig{figure=fig7Rcoluse.ps,height=3.5in}}

\figcaption{Source reconstructions at H (top-left), I (top-right), R (bottom-right) and 8~GHz. The
  I and R reconstructions used the smoothed images. The contours show the caustics of the
  lens model. }

\clearpage
\centerline{\psfig{figure=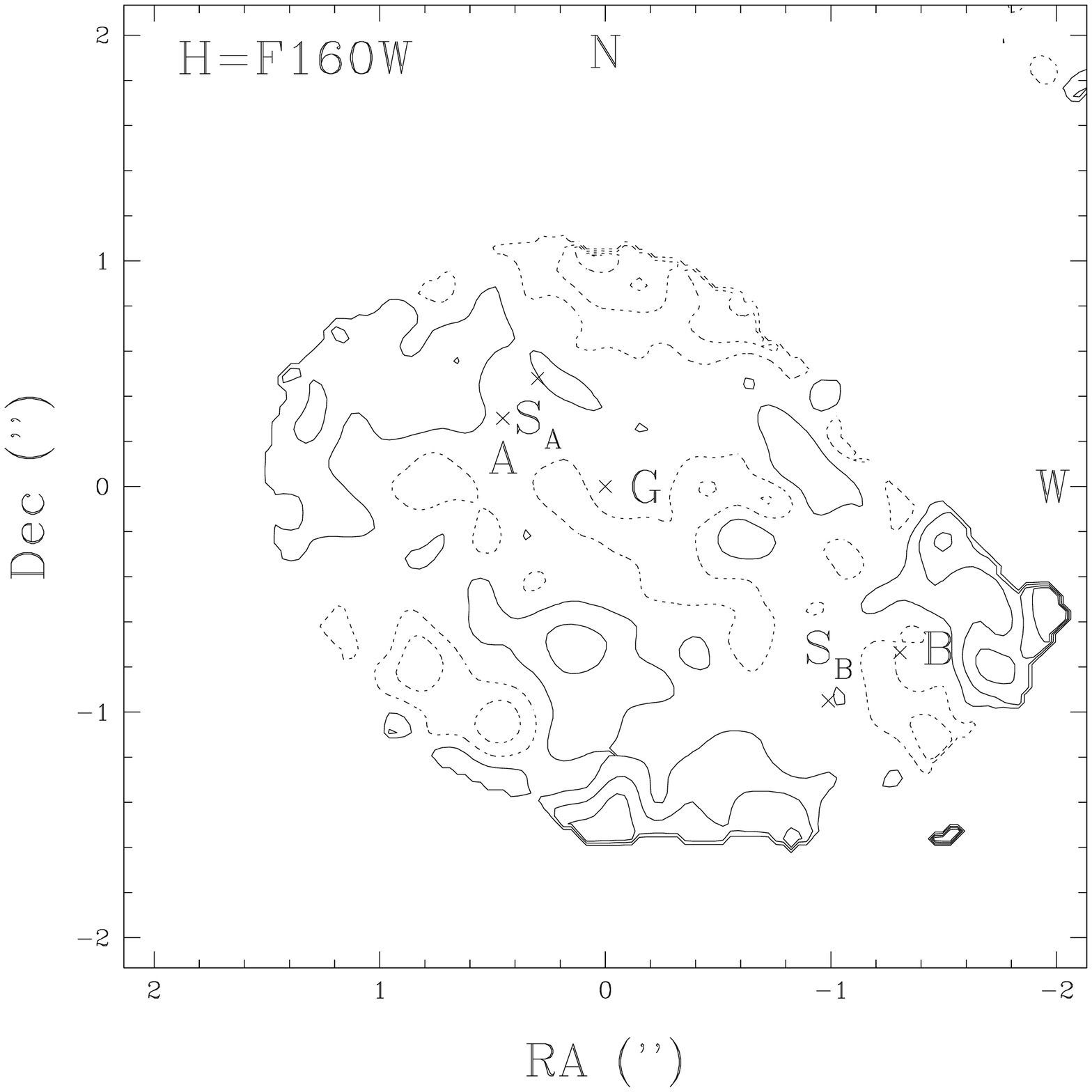,height=4.0in}}

\figcaption{Fractional ring residuals over the region with surface brightness 
 exceeding 10\% of the peak surface brightness.  The differential extinction 
 in the lens galaxy must satisfy $\Delta E(B-V) \leq 1.2 f$.  The solid contours 
 are drawn at $f=5\%$, $15\%$, $25\%$, and $35\%$, and the dashed contours are 
 drawn at $f=-5\%$, $-15\%$, $-25\%$, and $-35\%$. Most of the infrared ring, 
 including the regions corresponding to the gaps in the optical ring lie in the
  $|f| < 5\%$ region. }

\clearpage
  
\begin{deluxetable}{crrccccl}
\scriptsize
\tablecaption{Objects Near \MG}
\tablewidth{0pt}
\tablehead{ ID  &RA('') &Dec ('') &I (mag) &R--I (mag) &I--H (mag) &$\gamma_T$ &Comments  } 
\startdata
  G1   &   -0.5 &  -11.7 & $18.80\pm0.01$ &  $0.63\pm0.01$ &                &  0.113 & \nl
  G2   &   13.9 &    7.2 & $20.13\pm0.01$ &  $0.63\pm0.01$ &  $1.93\pm0.01$ &  0.046 & \nl
  G    &      0 &      0 & $21.04\pm0.01$ &  $1.26\pm0.02$ &  $2.81\pm0.01$ &        &  Lens Galaxy \nl
  S1   &   -7.2 &    6.6 & $21.87\pm0.01$ &  $1.15\pm0.01$ &  $(1.5\pm0.4)$ &        &  Star, A--6 \nl
  G3   &   -7.3 &   13.8 & $22.06\pm0.01$ &  $1.28\pm0.02$ &  $(2.9\pm0.4)$ &  0.019 &  Lens Group, A--5 \nl
  G4   &   -0.9 &   -6.0 & $22.25\pm0.02$ &  $0.48\pm0.02$ &  $1.40\pm0.02$ &  0.045 &  A--8 \nl
  G5   &    7.6 &   13.0 & $22.35\pm0.01$ &  $1.29\pm0.03$ &  $2.41\pm0.01$ &  0.017 &  Lens Group, A--3 \nl
  S2   &  -10.9 &  -19.7 & $22.45\pm0.01$ &  $1.36\pm0.02$ &                &        &  Star \nl
  G6   &   12.7 &    5.7 & $22.47\pm0.03$ &  $0.62\pm0.05$ &  $1.53\pm0.04$ &  0.018 & \nl
  G7   &   11.6 &   -3.4 & $22.59\pm0.03$ &  $1.19\pm0.05$ &  $2.52\pm0.02$ &  0.019 &  Lens Group, A--1 \nl
  G8   &    7.1 &    4.0 & $22.66\pm0.02$ &  $1.30\pm0.03$ &  $2.07\pm0.02$ &  0.028 &  Lens Group, A--9 \nl
  G9   &   -1.9 &    2.0 & $22.96\pm0.03$ &  $1.28\pm0.05$ &  $2.27\pm0.02$ &  0.071 &  Lens Group, A--D \nl
  G10  &   -9.7 &   -2.5 & $23.11\pm0.04$ &  $0.36\pm0.04$ &  $(3.0\pm0.4)$ &  0.018 &  A--7 \nl
  G11  &    4.2 &   20.4 & $23.15\pm0.03$ &  $0.42\pm0.04$ &                &  0.009 &  \nl
  G12  &    3.0 &   19.2 & $23.35\pm0.03$ &  $0.09\pm0.03$ &                &  0.008 &  \nl
  G13  &    2.0 &  -12.8 & $23.36\pm0.05$ &  $0.28\pm0.07$ &                &  0.013 &  \nl
  G14  &  -18.0 &    2.5 & $23.60\pm0.03$ &  $0.25\pm0.03$ &                &  0.008 &  \nl
  G15  &    0.2 &   -3.3 & $23.65\pm0.04$ &  $1.04\pm0.06$ &  $3.59\pm0.03$ &  0.043 &  Lens Group, A--C \nl
  G16  &    8.9 &    3.5 & $23.72\pm0.04$ &  $0.47\pm0.04$ &  $1.04\pm0.06$ &  0.014 &  \nl
  G17  &   -1.1 &  -14.6 & $23.72\pm0.05$ &  $0.58\pm0.09$ &                &  0.009 &  \nl
  G18  &  -15.0 &   -7.7 & $23.75\pm0.04$ &  $0.29\pm0.05$ &                &  0.008 &  \nl
  G19  &   -5.9 &  -12.8 & $23.80\pm0.04$ &  $0.36\pm0.04$ &                &  0.009 &  \nl
  G20  &   12.7 &   -4.0 & $23.85\pm0.06$ &  $0.22\pm0.06$ &  $1.87\pm0.06$ &  0.010 &  \nl
  G21  &    9.0 &  -11.6 & $23.90\pm0.06$ &  $0.93\pm0.11$ &                &  0.009 &  \nl
  G22  &    4.1 &   10.7 & $23.93\pm0.04$ &  $0.60\pm0.05$ &  $1.64\pm0.05$ &  0.011 &  \nl
  G23  &    2.7 &   18.2 & $24.03\pm0.05$ &  $0.24\pm0.04$ &                &  0.006 &  \nl
  G24  &   -5.6 &   15.7 & $24.18\pm0.05$ &  $0.18\pm0.05$ &                &  0.007 &  \nl
  G25  &   12.3 &   -2.2 & $24.42\pm0.06$ &  $0.31\pm0.07$ &  $1.78\pm0.07$ &  0.008 &  \nl
  G26  &  -11.2 &    9.6 & $24.46\pm0.07$ &  $1.13\pm0.11$ &                &  0.007 &  \nl
  G27  &  -15.2 &   -1.2 & $24.59\pm0.07$ &  $0.19\pm0.07$ &                &  0.006 &  \nl
  G28  &  -12.8 &   -1.6 & $24.66\pm0.08$ &  $0.27\pm0.09$ &                &  0.007 &  \nl
  G29  &    4.7 &   -4.5 & $24.72\pm0.09$ &  $0.51\pm0.11$ &  $1.67\pm0.14$ &  0.013 &  \nl
  G30  &  -12.9 &   -3.0 & $24.82\pm0.07$ &  $0.25\pm0.07$ &                &  0.006 &  \nl
  G31  &    5.5 &   12.3 & $24.97\pm0.09$ &  $0.91\pm0.15$ &  $4.40\pm0.08$ &  0.006 &  A--4 \nl
  G32  &    7.3 &   21.0 & $25.01\pm0.10$ &  $0.90\pm0.13$ &                &  0.003 &  \nl
  G33  &    6.7 &   -1.0 & $25.01\pm0.09$ &  $0.59\pm0.13$ &  $1.49\pm0.14$ &  0.011 &  \nl
  G34  &   10.8 &   -3.5 & $25.20\pm0.13$ &  $0.25\pm0.15$ &        $<1.77$ &  0.006 &  \nl
  \tablebreak
  G35  &   -2.3 &   -2.1 & $25.36\pm0.12$ &  $0.21\pm0.14$ &  $2.43\pm0.12$ &  0.021 &  \nl
  G36  &  -18.8 &    4.7 & $25.36\pm0.12$ &  $0.26\pm0.11$ &                &  0.003 &  \nl
  G37  &  -18.6 &    9.5 & $25.37\pm0.12$ &  $0.22\pm0.16$ &                &  0.003 &  \nl
  G38  &    4.0 &   -8.7 & $25.50\pm0.13$ &  $0.64\pm0.16$ &        $<1.65$ &  0.006 &  \nl
  G39  &  -12.9 &    9.5 & $25.51\pm0.12$ &  $0.46\pm0.13$ &                &  0.004 &  \nl
  G40  &   -1.8 &   -5.0 & $25.58\pm0.10$ & $-0.07\pm0.17\hphantom{-}$ & $-0.09\pm0.81$\hphantom{-} &  0.011 &  \nl
  G41  &    3.5 &  -13.6 & $25.60\pm0.14$ &  $0.16\pm0.15$ &                &  0.004 &  \nl
  S3   &   -2.6 &    7.8 & $25.60\pm0.11$ &  $0.18\pm0.09$ &  $2.21\pm0.10$ &        &  Star \nl
  G42  &   15.9 &   -5.9 & $25.88\pm0.12$ &  $0.10\pm0.15$ &        $<1.83$ &  0.003 &  \nl
\enddata
\tablecomments{SExtractor (Bertin \& Arnouts 1997) magnitudes and colors numbered in order of decreasing I 
  magnitude.  The primary lens galaxy is labeled G, the other galaxies and the stars in the PC field are 
  labeled G1, G2 $\cdots$ and S1, S2 $\cdots$ respectively in order of decreasing I magnitude The lens galaxy, 
  probable lens group members and stars are marked.  The labels A--? give the label of the object from
  Annis (1992).  Where we lack an H magnitude and Annis (1992) measured a K' magnitude, we include an estimated 
  I--H magnitude in parentheses using the mean H--K' magnitude for the objects we have in common.  
   The parameters for the primary lens G are the
  estimates from SExtractor, which may be biased by the ring.  For comparison the fitted values from \S2.4 are
  I$=$21.25 mag, R--I$=$1.28, and I--H$=$2.68 mag respectively.   The $\gamma_T$ column is the estimated external
  shear the galaxy would add to the lens model assuming the galaxies are singular isothermal spheres at the same
  redshift obeying the Faber-Jackson (1976) relation (see \S2.5). 
  }  
\end{deluxetable}

\end{document}